\documentclass{lmcs} 
\pdfoutput=1

\usepackage{lastpage}
\lmcsdoi{21}{3}{3}
\lmcsheading{}{\pageref{LastPage}}{}{}%
{May~03,~2024}{Jul.~15,~2025}{}

\usepackage[utf8]{inputenc}


\usepackage{hyperref}
\usepackage{amsmath,amsthm,amssymb, amsbsy}
\usepackage{caption}
\usepackage{subcaption}
\usepackage[capitalize]{cleveref}
\usepackage{tikz} 
\usetikzlibrary{automata, arrows, positioning}
\usepackage{thm-restate}
\usepackage{tikz}

\theoremstyle{remark}
\newtheorem{case}{Case}
\newtheorem{subcase}{Sub-case}[case]

\usepackage{etoolbox}
\AtBeginEnvironment{proof}{\setcounter{case}{0}}
\AtBeginEnvironment{proof}{\setcounter{subcase}{0}}
\AtBeginEnvironment{claimproof}{\setcounter{case}{0}}

\usepackage{xcolor}

\newcommand{\N}{\mathbb{N}}
\newcommand{\Nmax}{\N_{\max}}
\newcommand{\matrices}[2]{\mathcal{M}^{#1 \times #2}}
\newcommand{\matricesOm}[2]{\mathcal{M}^{#1 \times #2}_\Omega}
\newcommand{\ideal}{\mathfrak{M}}
\newcommand{\autA}{\mathcal{A}}
\newcommand{\autB}{\mathcal{B}}
\newcommand{\brfunc}[1]{[\![#1]\!]}

\newcommand{\funcA}{\brfunc{\autA}}
\newcommand{\funcB}{\brfunc{\autB}}

\newcommand{\matricesOmbar}[2]{\mathcal{M}_{\overline{\Omega}}^{#1\times #2}}

\newcommand{\aval}{\mathsf{val}_\autA}
\newcommand{\bval}{\mathsf{Val}_\autB}

\newcommand{\nopath}{-}

\newenvironment{smallpmatrix}
  {\left(\begin{smallmatrix}}
  {\end{smallmatrix}\right)}
  
\newcommand{\monoidid}{\mathsf{i\kern-0.8pt d}}

\newcommand{\ceil}[1]{\left\lceil#1\right\rceil}

\usepackage{enumitem}

\theoremstyle{remark}
\crefname{runningexample/}{Running Example, Part}{Running Example, Parts}
\newtheorem{runningexample/}{Running Example, Part}

\newenvironment{runningexample}
  {%
   \pushQED{\qed}\begin{runningexample/}}
  {\popQED\end{runningexample/}}

\crefname{defi}{Definition}{Definitions} %
\crefname{thm}{Theorem}{Theorems}
\crefname{rem}{Remark}{Remarks}
\crefname{lem}{Lemma}{Lemmas} 
\crefname{prop}{Proposition}{Propositions}

\begin{document}

\title{The Big-O Problem for Max-Plus Automata is Decidable (PSPACE-Complete)}
\thanks{The first author has been supported for this work by the
EPSRC grant EP/T018313/1. The second author was partially supported by the ERC grant INFSYS, agreement no. 950398.
A preliminary version appeared in \cite{DaviaudP23}. 
}

\author[L.~Daviaud]{Laure Daviaud\lmcsorcid{0000-0002-9220-7118}}[a]
\author[D.~Purser]{David Purser\lmcsorcid{0000-0003-0394-1634}}[b]
\author[M.~Tcheng]{Marie Tcheng}[c]

\address{University of East Anglia, UK}
\email{l.daviaud@uea.ac.uk} 

\address{University of Liverpool, UK}
\email{d.purser@liverpool.ac.uk} 

\address{ENS Paris Saclay, France}
\email{marie.tcheng@ens-paris-saclay.fr} 

\begin{abstract}
We show that the big-O problem for max-plus automata, i.e. weighted automata over the semiring  $(\mathbb{N}\cup \{-\infty\}, \max, +)$, is decidable and PSPACE-complete. The big-O (or affine domination) problem asks whether, given two max-plus automata computing functions $f$ and $g$,  there exists a constant $c$ such that $f \leq cg+ c$. This is a relaxation of the containment problem asking whether $f \le g$, which is undecidable. 
Our decidability result uses Simon's forest factorisation theorem, and relies on detecting specific elements, that we call witnesses, in a finite semigroup closed under two special operations: stabilisation and flattening.
\end{abstract}

\maketitle

\section{Introduction}

Weighted automata are a generalisation of finite state automata, assigning values (integers, rationals, strings...) to input words, and modelling probabilities, costs or program running times. They have been introduced by Sch\"utzenberger \cite{Schutzenberger61b} and found applications in quantitative verification \cite{ChatterjeeDH10} and verification of probabilistic systems \cite{Vardi85}, text and speech recognition \cite{MohriPR02} or program complexity analysis  \cite{ColcombetDZ14}. 

More precisely, a weighted automaton is defined over a semiring. Commonly studied examples include the rational semiring $(\mathbb{Q},+,\times)$ (and the particular case of probabilistic automata) and the tropical semirings $(\mathbb{N}\cup\{+\infty\},\min,+)$ (referred to as min-plus automata) and  $(\mathbb{N}\cup\{-\infty\},\max,+)$ (max-plus automata). Non-deterministic finite automata can be viewed as weighted automata over the Boolean semiring. In all these cases, input words are mapped to rational values (and possibly $+\infty$ or $-\infty$).

Comparing the functions computed by two given weighted automata is then a fundamental question. It is natural to consider the equivalence problem (are they equal?), and the containment problem (is one pointwise smaller than the other?). These two problems have been extensively studied and solutions are highly dependent on the semiring under consideration. Results for the equivalence problem are contrasting, but the containment problem is usually difficult:
\begin{itemize}
    \item For the Boolean semiring, the equivalence and containment problems correspond to the language equivalence and language inclusion respectively. Both problems are PSPACE-complete~\cite{StockmeyerM73}.
    \item For the rational semiring the equivalence problem is decidable~\cite{Schutzenberger61b}, even in polynomial time \cite{kiefer13}, but containment is undecidable~\cite{paz2014introduction}, even in restricted subclasses~\cite{DaviaudJLMP021}.
    \item For the tropical semirings both problems are undecidable~\cite{Krob94}. See~\cite{AlmagorBK22,AlmagorBK11} for a comprehensive overview of the decidability boundary for min-plus automata.
\end{itemize}

In this paper we consider a relaxation of the containment problem, called the big-O problem which asks whether an automaton $\autA$ is big-O of an automaton $\autB$, that is, if there exists a constant $c$ such that:
\[
\funcA(w) \le c\funcB(w) + c\text{ for all words } w
\]
where $\funcA$ (resp. $\funcB$) denotes the function computed by $\autA$ (resp. $\autB$).
Intuitively the problem asks whether, asymptotically, $\funcB$ grows at least as fast as $\funcA$ on every sequence of words.

Chistikov, Kiefer, Murawski and Purser~\cite{ChistikovKMP20,ChistikovKMP22} study the big-O problem over the non-negative rational semiring, where it is shown to be undecidable in general, but decidable for certain restrictions on the ambiguity or the accepted language. Similarly, in~\cite{CzerwinskiLMPW22}, two restrictions of the big-O problem, also known to be undecidable in general, are studied on specific subclasses: the boundedness problem (a.k.a.~limitedness), and the zero isolation problem.\footnote{The boundedness problem is the special case of the big-O problem when $\funcB = 1$. The zero isolation problem is the special case when $\funcA = 1$. Chistikov et al.~\cite{ChistikovKMP22} deal with the problem whether $\funcA \le c\funcB$ - so without $+c$, but the two problems are equivalent (see \cref{remark:bigowithoutplusc}).}

For the tropical semirings, the big-O problem has also been proved to be decidable in the $(\mathbb{N}\cup\{+\infty\},\min,+)$ setting via the study of another problem: the domination problem~\cite{Colcombet07, Colcombet09}. This later asks whether there is a function $\alpha:\mathbb{N} \to \mathbb{N}$ such that  $\funcA \le \alpha \circ \funcB$. Affine domination requires that $\alpha$ be affine, and is equivalent to the big-O problem that we consider. Colcombet and Daviaud~\cite{ColcombetD13} show that domination and affine domination are equivalent and decidable for min-plus automata. More specifically, it turns out that if some function $\alpha$ exists then an affine $\alpha$ suffices. This result superseded the decidability of the boundedness problem for min-plus automata~\cite{Hashiguchi91a, leung1988topological, Simon94}.

In this paper, we turn our attention to $(\mathbb{N}\cup\{-\infty\},\max,+)$ for which the (un)decidability of the big-O problem was open. First, note that there is no obvious way to use the results obtained for min-plus automata. The natural transformation - given $f$ computed by a min-plus automaton, $-f$ is computed by a max-plus automaton - does not preserve positivity, and the standard way to go back to $\mathbb{N}$ implies adding a big enough function to $-f$ which does not preserve the growth rate. In fact, we prove that the equivalence between domination and affine domination does not hold any more for max-plus automata (see Running Example~\ref{re:notbig-O}). Second, the boundedness problem in this case is trivially decidable and does not provide any help. The problem for max-plus automata requires individual attention and the introduction of new tools. We show that it is PSPACE-complete and our proof provides new insights in the description of the behaviour of these automata. In~\cite{ColcombetDZ14}, some description of the asymptotic behaviour of the functions computed by max-plus automata is given, but this is somehow orthogonal to the big-O problem. While providing a precise description, it is not sufficient to solve the big-O problem and new techniques are required.

Building on some standard techniques, in particular Simon's factorisation forest theorem and the stabilisation operation~\cite{Simon90,Colcombet07,Kufleitner08,KirstenL09,FijalkowGKO15,ColcombetD13,CzerwinskiLMPW22}, we construct a finite semigroup closed under the stabilisation operation and a new \emph{flattening} operation. The stabilisation operation identifies unbounded behaviour, while the flattening operation identifies maximal growth rates. The problem reduces to detecting the presence of \emph{witnesses} in this semigroup. A naïve search through the semigroup gives decidability, but may require exponential space. The PSPACE complexity comes from searching witnesses of a particular shape only requiring polynomial space. The hardness comes from the PSPACE-hardness of the universality problem for Boolean automata.

\paragraph*{Organisation of the paper} The paper expands on \cite{DaviaudP23}, giving full proofs and explanations. In Section~\ref{sec:mpa}, we give the definition of max-plus automata and introduce a running example that we will use all along the paper. In Section~\ref{sec:bigO}, we state the big-O problem and a simplified version of it, prove their PSPACE-hardness and reduce our result to prove that the simplified big-O problem is in PSPACE (Theorem~\ref{thm:main-simplified}). In Section~\ref{section:projection}, we give a high-level description of this proof and define the semigroups that will be used and the stabilisation and flattening operations. In Section~\ref{sec:decision}, we define witnesses and give a decision procedure that we show to be PSPACE. Sections~\ref{section:facto-fault}, \ref{1implies4}, \ref{4implies3} and~\ref{2implies1} are then dedicated to prove that the decision procedure is sound and complete. The organisation and content of these sections are explained at the end of Section~\ref{sec:decision}, when the suitable notions have been introduced. Finally, in Section~\ref{sec:example}, we provide a sequence of max-plus automata that require increasingly intricate witnesses (for a notion defined in the paper); this section is new compared with \cite{DaviaudP23}.

\section{Max-plus automata}
\label{sec:mpa}

Let $\Nmax$ denote the set $\N \cup \{-\infty\}$ and note that $(\Nmax, \max, +)$ is a semiring. For some positive integers $i,j$, let $\matrices{i}{j}$ denote the set of matrices of dimension $i \times j$ with coefficients in $\Nmax$. We define the product of matrices as usual on a semiring, provided the numbers of columns of $A$ matches the number of rows of $B$: $(A\otimes B)_{q,q'} = \max_{q''} \left(A_{q,q''} + B_{q'',q'}\right)$. We will use the symbol $\otimes$ to denote the product of matrices on several semirings, but the context will always clearly identify which one.
\begin{defi}
A \emph{max-plus automaton} is a tuple $\langle Q, \Sigma, M, I, F\rangle$ where $Q$ is a finite set of states (and $|Q|$ denotes the number of states), $\Sigma$ is a finite alphabet, $M: \Sigma \to \matrices{|Q|}{|Q|}$ maps each letter to a matrix, $I$ is a row vector in $\matrices{1}{|Q|}$ and $F$ a column vector in $\matrices{|Q|}{1}$. Moreover, the automaton is said to be deterministic if $I$ has at most one entry different from $-\infty$ and for all $a$ in $\Sigma$, every row of $M(a)$ has at most one entry different from $-\infty$.
\end{defi}

Given a max-plus automaton $\langle Q, \Sigma, M, I, F\rangle$, we extend $M$ by morphism to $\Sigma^*$.

\begin{defi}
The \emph{weighting function} computed by $\autA = \langle Q, \Sigma, M, I, F\rangle$, a max-plus automaton, is defined as the function $\funcA:\Sigma^*\to \Nmax$ mapping a word $w = w_1w_2\dots w_k$, where $w_i \in \Sigma$ for all $i=1,\ldots,k$, to:
\[
\funcA(w) = I \otimes M(w_1)\otimes M(w_2)\otimes \dots \otimes M(w_k) \otimes F.
\]
\end{defi}

These definitions can be expressed in terms of graphs as usual, and we will rather use the usual automaton vocabulary (transitions, runs, accepting runs, initial and final states, etc.) when appropriate in some proofs. We will often write $p\xrightarrow{w:x}q$ for a run from state $p$ to state $q$ labelled by the word $w$ with weight $x \neq -\infty$, the weight of a run being the sum of the weights of the transitions in the run. In matrix terms, this means that, for $w = w_1w_2\dots w_k$, where $w_i \in \Sigma$ for all $i=1,\ldots,k$, there are $p=j_0, \ldots, j_k=q$ such that $M(w_i)_{j_{i-1},j_i} = x_i$ and $x = x_1 + \ldots + x_k$. The run is accepting if $p$ is initial and $q$ is final, i.e. $I_p \neq -\infty$ and $F_q \neq -\infty$ and $\funcA(w)$ is equal to the maximum of the weights of the accepting runs labelled by $w$.

We assume that all the states in the automata under consideration in this paper are accessible and co-accessible.

The size of an automaton is the number of bits required to encode $M$, $I$ and $F$ which is bounded by $(|\Sigma|\cdot |Q|^2 +2|Q|) \cdot \ceil{\log(\Lambda)}$, where $\Lambda$ is the maximal weight appearing in an entry of $M, I $ or $F$. 

Throughout the paper we will illustrate the results and proofs using a running example that we detail now.

\begin{figure}
     \centering
     \begin{subfigure}[b]{0.3\linewidth}
         \centering
         \includegraphics[width=0.7\textwidth]{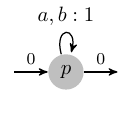}
         \caption*{$\autA$: Computes word length $|w|$.\\ \ }
         
     \end{subfigure}
     \hfill
     \begin{subfigure}[b]{0.6\linewidth}
         \centering
         \includegraphics[width=0.9\textwidth]{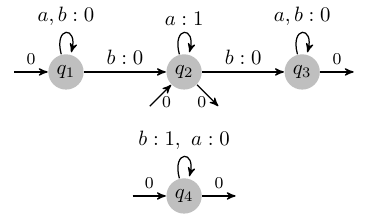}
         \caption*{$\autB$: Computes the maximum between the longest block of $a$'s and the number of $b$'s.  }
     \end{subfigure}
        \caption{Running examples}
        \label{fig:running}
\end{figure}

\begin{runningexample}\label{eg:running}
Let us consider two example automata $\autA$ and $\autB$ over $\Sigma = \{a,b\}$, depicted in \cref{fig:running}.
$\autA$ computes the length of the input word and $\autB$ computes the maximum between the length of the longest block of consecutive $a$'s, and the number of $b$'s.

In the following matrix descriptions, to avoid cluttering notation, we use $\nopath$ instead of $-\infty$ to denote that there is no path.

Formally, the automaton $\autA$ is defined by a single state $Q_\autA = \{p\}$, alphabet $\Sigma = \{a,b\}$ and $M_{\autA}(a) = (1), M_{\autA}(b) = (1)$ with $I_\autA = F_\autA = (0)$. 

Formally $\autB$ is defined by states $Q_\autB = \{q_1,q_2,q_3,q_4\}$, alphabet $\Sigma = \{a,b\}$ and 
\[M_{\autB}(a) = 
\begin{smallpmatrix}
0 & \nopath & \nopath & \nopath\\
\nopath & 1 & \nopath & \nopath\\
\nopath & \nopath & 0 & \nopath\\
\nopath & \nopath & \nopath & 0
\end{smallpmatrix} \text{ and }M_{\autB}(b) = 
\begin{smallpmatrix}
0 & 0 & \nopath & \nopath\\
\nopath & \nopath & 0 & \nopath\\
\nopath & \nopath & 0 & \nopath\\
\nopath & \nopath & \nopath & 1
\end{smallpmatrix}\]
with $(I_\autB)_{q_1} =(I_\autB)_{q_2} = (I_\autB)_{q_4} =(F_\autB)_{q_2} = (F_\autB)_{q_3} = (F_\autB)_{q_4} = 0$, and the unspecified entries of $I,F$ are $-\infty$. 
\end{runningexample}

\section{Decidability of the big-O problem}
\label{sec:bigO}

The big-O problem for max-plus automata asks whether, given two max-plus automata $\autA,\autB$ on the same alphabet $\Sigma$, there is a positive integer $c$ such that for all words $w$ in $\Sigma^*$, $\funcA(w) \leq c\funcB(w) + c$. In this case, we say that $\autA$ is big-O of $\autB$.
\begin{runningexample}\label{re:notbig-O}

Observe that $\autA$ is not big-O of $\autB$, since $\funcA((a^{n-1}b)^n) = n^2$, while $\funcB((a^{n-1}b)^n) = n$ for all positive integers $n$.

However, $\funcB(w) \le \funcA(w)$ for all words $w$, so $\autB$ is big-O of $\autA$.

Note also that $\funcA \le (\funcB + 1)\funcB$, hence $\funcA$ is dominated by $\funcB$ as explained in the introduction, but not big-O, showing the discrepancy between the min-plus and the max-plus cases. 
\end{runningexample}

The main contribution of this paper is the following result.

\begin{thm}\label{thm:main}
The big-O problem for max-plus automata is decidable and is PSPACE-complete.
\end{thm}

The rest of this paper is devoted to prove this theorem. The first step is to make a number of simplifications on the automata taken as input.

\begin{thm}\label{thm:main-simplified}
The following problem, called the \emph{simplified big-O problem}, is PSPACE-complete: 
\begin{itemize}
\item Input: Max-plus automata $\autA,\autB$ such that $\autA$ is deterministic and $\funcB:\Sigma^* \to \N$.
\item Output: Yes if and only if $\autA$ is big-O of $\autB$.
\end{itemize}
\end{thm}
Compared to the big-O problem, the simplified big-O problem requires that $\autA$ be deterministic and no word $w$ has $\funcB(w) = -\infty$. 

\begin{runningexample}
Recall $\autA,\autB$ from the running example in \cref{eg:running}. Observe that $\autA$ is deterministic and $\funcB(w) \ge 0$ for all $w$, so $\autA$ and $\autB$ form an instance of the simplified big-O problem.
\end{runningexample}

\begin{prop}\label{prop:tosimplified}
If the simplified big-O problem is decidable in PSPACE then the big-O problem is decidable in PSPACE.
\end{prop}

\begin{proof}
We reduce an instance of the big-O problem to the simplified big-O problem. The instance of simplified big-O problem will be of polynomial size with respect to the initial input, but a PSPACE pre-processing step is also used.
\begin{itemize}
\item Let $\autA,\autB$ be max-plus automata. Let us first construct $\autA',\autB'$ with $\brfunc{\autB'}:\Sigma^* \to \N$ such that $\autA$ is big-O of $\autB$ if and only if  $\autA'$ is big-O of $\autB'$. Let $L_\autA$ (resp. $L_\autB$) be the (rational) language of words $w$ such that $\funcA(w) \neq -\infty$ (resp. $\funcB(w) \neq -\infty$). Checking whether $L_\autA \subseteq L_\autB$ can be done in PSPACE (consider the Boolean automata obtained from $\autA$ and $\autB$ by ignoring the weights - they accept $L_\autA$ and $L_\autB$, and inclusion of rational languages is PSPACE~\cite{MeyerS72}). If $L_\autA$ is not included in $L_\autB$ then $\autA$ cannot be big-O of $\autB$. Take $\autA'$ that computes the length of the words and $\autB'$ that computes the function $0$. If $L_\autA$ is included in $L_\autB$, take $\autA' = \autA$ and $\autB'$ being $\autB$ augmented with a new state which is both initial and finial and has a self loop on all letters with weight $0$. In both cases, $\autA$ is big-O of $\autB$ if and only if  $\autA'$ is big-O of $\autB'$ and $\brfunc{\autB'}:\Sigma^* \to \N$.

\item We now reduce this to the case where automaton $\autA$ is deterministic. Consider two automata $\autA = \langle Q_\autA, \Sigma, M_\autA, I_\autA, F_\autA\rangle$ and $\autB = \langle Q_\autB, \Sigma, M_\autB, I_\autB, F_\autB\rangle$ with $\brfunc{\autB}:\Sigma^* \to \N$. We construct $\autA',\autB'$ with $\autA'$ deterministic and $\brfunc{\autB'}:\Sigma^* \to \N$ such that $\autA$ is big-O of $\autB$ if and only if $\autA'$ is big-O of $\autB'$. Automata  $\autA'$ and $\autB'$ are over the alphabet $\Sigma' = \{a_q \mid a \in \Sigma\text{, } q \in Q_\autA \}$ and: 
\begin{itemize}
\item $\autA'$ is constructed from $\autA$, with states $Q_\autA \cup \{r\}$ with $r$ a new state (the new unique initial state), $(I_{\autA'})_r = 0$ and all the other entries of $I_{\autA'}$ being $-\infty$, final states $(F_{\autA'})_p = (F_{\autA})_p$ for $p\in Q_\autA$ and $(F_{\autA'})_r = \max\{(F_\autA)_p \mid p \text{ initial in $\autA$ }\}$ , a transition $p\xrightarrow{a_q:x}q$ for each transition $p\xrightarrow{a:x}q$ in $\autA$, and a transition $r\xrightarrow{a_q: x+(I_\autA)_p}q$ for each transition $p\xrightarrow{a:x}q$ in $\autA$. Note that $\autA'$ is deterministic.
\item  $\autB'$ is constructed from $\autB$, with states $Q_\autB$, initial and final states $I_\autB$ and $F_\autB$ respectively, and a transition $p\xrightarrow{a_r:x}q$ for each transition $p\xrightarrow{a:x}q$ in $\autB$ and each $r \in Q_\autA$. Note that $\brfunc{\autB}:\Sigma^* \to \N$.
\end{itemize}
Suppose that $\autA$ is big-O of $\autB$ and let us prove that $\autA'$ is big-O of $\autB'$. Let $w$ be a word over $\Sigma'$ and $\bar{w}$ over $\Sigma$ defined as $w$ where the subscripts of the letters are removed. Then $\brfunc{\autA'}(w) \leq \funcA(\bar{w})$ and $\funcB(\bar{w}) = \brfunc{\autB'}(w)$ by construction and hence $\autA'$ is big-O of $\autB'$. Conversely, suppose that $\autA'$ is big-O of $\autB'$. Let $w \in \Sigma^*$. Consider an accepting run in $\autA$ labelled by $w$ that has maximal weight and $q_0, q_1, q_2, \ldots, q_k$ the corresponding sequence of states. Let $\bar{w} = a^1_{q_1} \dotsm a^k_{q_k}$ word over $\Sigma'$ where $w = a^1 \dotsm a^k$ and $a^i \in \Sigma$ for all $i$. Then $\funcA(w) = \brfunc{\autA'}(\bar{w})$ and $\brfunc{\autB'}(\bar{w}) = \funcB(w)$ by construction. Hence $\autA$ is big-O of $\autB$.\qedhere
\end{itemize}
\end{proof}

\begin{prop}\label{prop:hardness}
The simplified big-O problem is PSPACE-hard.
\end{prop}

\begin{proof}
We reduce from the  \textsc{Cofiniteness} problem, which asks whether the language of a non-deterministic finite automaton is cofinite, that is, accepts all but a finite set of words. 

\begin{lem}\label{prop:cofinitenesshardness}
\textsc{Cofiniteness} is PSPACE-hard 
\end{lem}
\begin{proof}
We reduce from the universality of non-deterministic finite automata. Let $\autA$ be an instance of universality over alphabet $\Sigma$; we construct $\autA'$ over $\Sigma\cup\{\#\}$ where $\#$ is a new letter not in $\Sigma$. $\autA'$ is such that for all words $w$ over $\Sigma^*$ and for all words $u$ over $\Sigma\cup\{\#\}$,   $w\#u$ is accepted by $\autA'$ if and only if $w$ is accepted in $\autA$. This is achieved by augmenting $\autA$ with an accepting sink state reachable on $\#$ from every accepting state of $\autA$. We observe $\autA$ is universal if and only if $\autA'$ is cofinite. In particular, if $\autA$ is universal, so too is $\autA'$, and in particular, cofinite. If $\autA$ does not accept $w$ then $\autA$ does not accept the language $w\#\Sigma^*$, and thus is not cofinite.
\end{proof}

Let $\autB$ be an input to \textsc{Cofiniteness}; we construct an instance of the simplified big-O problem. Let $\autA'$ and $\autB'$ such that $\brfunc{\autA'}(w) = |w|$ for all $w\in\Sigma^*$ and $\brfunc{\autB'}(w) = |w|$ for all $w$ accepted by $L_\autB$ and 0 otherwise. The automaton $\autB'$ is constructed from $\autB$ by associating every edge in $\autB$ with the weight $1$ and augmented with a new state which is both initial and finial and has a self loop on zero for all letters.

We observe $\autB$ is cofinite if and only if $\autA$ is big-O of $\autB$.
\begin{itemize}
    \item If $\autB$ is cofinite then there exists a longest word $w_L$ that is not recognised by $\autB$ and $\brfunc{\autA'}(w) \le \brfunc{\autB'}(w) + |w_L|$ for all $w\in\Sigma^*$ and $\autA$ is big-O of $\autB$. 
    \item If $\autB$ is not cofinite then there is an infinite sequence of words not accepted by $\autB$ and in particular one of increasing length words, $(w_i)_{i\in \mathbb{N}}$. Hence we have $\brfunc{\autA'}(w_i) \to \infty$ while $\brfunc{\autB'}(w_i) = 0$ as $i\to\infty$ and $\autA'$ is not big-O of $\autB'$.\qedhere
\end{itemize}
\end{proof}

\cref{prop:tosimplified} states that it is enough to prove that the simplified big-O problem is in PSPACE in order to conclude the proof of \cref{thm:main}. \cref{prop:hardness} gives the hardness parts of \cref{thm:main,thm:main-simplified} (since the simplified problem is a particular instance of the general one).
 
The rest of the paper, \cref{section:projection} and beyond, will then focus on proving that the simplified big-O problem is PSPACE.

\begin{rem}\label{remark:bigowithoutplusc}
Chistikov, Kiefer, Murawski and Purser~\cite{ChistikovKMP22} use a slightly different notion of big-O for rational automata, requiring the existence of $c$ such that $\funcA \le c \funcB$ (that is, without $+c$). We note that for max-plus automata these two problems are equivalent, they reduce to each other in PSPACE, so without further blow up in complexity.
Indeed, $\funcA \le c\funcB+c$ if and only if $\funcA \le c(\funcB+1)$, for which the translation can be constructed in polynomial time.
Conversely, there exists $c$ such that $\funcA \le c\funcB$ if and only if there is $c$ such that $\funcA \le c\funcB+ c$ and $\{w \mid \funcA(w) \ge 1 \text{ and } \funcB(w) = 0\}$ is empty. The latter check can be done with an emptiness test for regular languages in PSPACE.
\end{rem}
\begin{rem}
One could be interested in computing the optimal or minimal constant $c$ such that $\funcA \le c\funcB + c$, or with the previous remark such that  $\funcA \le c\funcB$. Such a $c$ is not computable. If one could compute it and check whether it is at most $1$, it could be decided whether $\funcA \le \funcB$, which is undecidable for max-plus automata~\cite{Krob94}.
\end{rem}

\section{Projective semigroups}
\label{section:projection}

From now on, we fix a deterministic max-plus automaton $\autA = \langle Q_\autA, \Sigma, M_\autA, I_\autA, F_\autA\rangle$ and a max-plus automaton $\autB = \langle Q_\autB, \Sigma, M_\autB, I_\autB, F_\autB\rangle$ over the same alphabet $\Sigma$.

\subsection{Proof schema}

We describe here the general idea of the proof: this is a very high level explanation, omitting a lot of technical (but necessary) details, so should be read as such.

We want to show that either $\autA$ is big-O of $\autB$, or exhibit an infinite sequence of words witnessing that this is not the case. The general idea of the proof follows the (now rather standard) scheme of finding a finite semigroup, computable from $\autA$ and $\autB$, and identifying special elements in it - which we will call witnesses - such that witnesses exist in the semigroup associated with $\autA$ and $\autB$ if and only if  $\autA$ is \emph{not} big-O of $\autB$.
To obtain the complexity bound, we will show that testing the existence of such a witness can be done in PSPACE.
Similar proof schema can be found in~\cite{Simon94,ColcombetD13,ColcombetDZ14} - in particular this has been used for deciding the boundedness of distance automata, and we will explain the new insights our proof provides in comparison.

To construct the appropriate semigroup, the first step is to project the weights into $\{-\infty, 0, 1\}$. Indeed, the exact weights are not important, the only thing that matters is the difference in growth rates as illustrated in \cref{re:notbig-O}. This leads to introduce the so-called semigroup of paths (denoted $\overline{\ideal}_{\autA,\autB}$) in~\cref{def:sgofp}, which is finite and contains elements $(p,x,q,M)$ where $p,q$ are states in $\autA$, $x$ is in $\{0, 1\}$ and there is a word $w$ such that there is a run in $\autA$ from $p$ to $q$ labelled by $w$ with weight $0$ if $x = 0$, and positive weight if $x =1$. Moreover, $M$ is the matrix $M_\autB(w)$ where all the (strictly) positive entries are replaced by $1$. This semigroup is easily computed starting from the letters and closing under an appropriate product.

This semigroup witnesses the existence of runs with $0$ or positive weights, but is not enough to compare the growth rates in $\autA$ and $\autB$. The next step is to add a value $\infty$: an entry will be $\infty$ if there is a sequence of words with paths with unbounded weights for this entry. We introduce the semigroup of asymptotic behaviour (denoted $\ideal_{\autA,\autB}$) in \cref{def:sgab}, which contains elements $(p,x,q,M)$ where $p,q$ are states in $\autA$, $x$ is in $\{0, 1, \infty\}$ and $M$ is a matrix with entries in $\{-\infty, 0, 1, \infty\}$. We introduce a first operation - the stabilisation operation (as defined in \cref{def:op:stab}) - which essentially iterates a word many times and puts value $\infty$ in the entries that are unbounded as the word is repeated. If starting from the letters and closing under an appropriate product and the stabilisation operation, we would get the following: for an element $(p,x,q,M)$ in the semigroup of asymptotic behaviour, there is a sequence of words $(w_i)_i$ labelling paths in $\autA$ from $p$ to $q$ with weights $0$ if $x=0$, positive but bounded weights if $x=1$, unbounded weights if $x=\infty$. Moreover, an entry of $M$ is $0$ if it is $0$ in all $M_\autB(w_i)$, is $1$ if it is positive and bounded for the $M_\autB(w_i)$ and $\infty$ if it tends to $+\infty$ when $i\to +\infty$ in the matrix $M_\autB(w_i)$.

Up to now the semigroup of asymptotic behaviour witnesses unboundedness of sequences of words, but not yet the difference in growth rates in $\autA$ and $\autB$. This is where our proof gives new insight. We introduce a second operation, which we call flattening, and behaves as follows: consider an element $(p,x,p,M)$ in the semigroup of asymptotic behaviour such that $x=\infty$, so witnessing some sequence of words with unbounded weights in $\autA$ from $p$ to $p$. We would like to iterate the words in this sequence, no longer looking at unboundedness, but rather at which entries in $M$ will grow linearly as the elements of the sequence are repeated. We define the flattening operation to do exactly that. After flattening, we obtain an element $(p,x,p,M')$ witnessing a sequence of words $(w_i)_i$ such that it is possible for an entry in $M'$ to have value $1$, corresponding to paths with unbounded weights, but only if the asymptotic growth rate of these weights is little-o of the respective weights in $\autA$ from $p$ to $p$. In other words, unbounded runs of maximal growth rate continue to be represented by $\infty$, but runs which are not of the maximal growth rate are `flattened' back to $1$, even if they are unbounded. 

The semigroup of asymptotic behaviour can be easily computed starting from the letters, closing under an appropriate product and stabilisation and flattening operations. Witnesses are those elements $(p,x,q,M)$ where $x = \infty$, with $p$ initial and $q$ final in $\autA$, and all the entries of $M$ between an initial and a final state in $\autB$ have value at most $1$. We will use Simon's factorisation theorem~\cite{Simon90} (see~\cref{theorem:Simon90}) in two ways. In the case where there is no witness, this will allow us to bound a constant $c$ such $\funcA \le c\funcB + c$. In the case where there is a witness, this will allow us to limit our search to a specific type - called tractable witness of non-domination (\cref{def:tractable_witness}) - ensuring the PSPACE complexity of our algorithm.

\begin{rem}
We could have introduced a single operation, somehow defined as a stabilisation in some cases, and as flattening in others. We chose not to as the definition felt ad-hoc and we believe the two operations of stabilisation and flattening are more intuitive. This might have slightly simplified parts of the proof, however an added benefit we gained with two operations is to characterise exactly the shape of these special witnesses we define: the tractable witnesses of non-domination.
\end{rem}
\pagebreak[5]
\subsection{Semigroup of paths}
\label{sec:semi-group-of-paths}
\paragraph*{Projection in $\{-\infty,0,1,\infty\}$}

Let $\Omega$ be the semiring $\left(\{-\infty,0,1,\infty\}, \max, +\right)$ where the operations are defined as follows: the $\max$ operation is given by the order $-\infty < 0 < 1 < \infty$ and the sum operation is commutative and given by $-\infty + x = -\infty$ for any element $x$ (including $x=\infty$), $0 + x = x$ for any $x \in \{0,1, \infty\}$, $1 + 1 = 1$ and $1 + \infty = \infty + \infty = \infty$.

Let $\matricesOm{i}{j}$ be the set of matrices of size $i \times j$ over the semiring $\Omega$. We use again $\otimes$ to denote the product of matrices induced by the operations in $\Omega$.

Given a finite set $Q$ and a positive integer $i$, we denote by $(\ideal_{Q,i}, \otimes)$ the semigroup where:
\begin{itemize}
\item $\ideal_{Q,i}$ is the union of the sets $Q \times \Omega \times Q \times \matricesOm{i}{i}$ and $\{\bot\}$.
\item $(p,x,q,M)\otimes(p',x',q',M') = (p,x + x',q', M\otimes M')$ if $q = p'$ and $\bot$ otherwise.
\end{itemize}

We will often denote the product of two elements $e,e'\in\ideal_{Q,i}$ as $ee'$ instead of $e\otimes e'$.

\paragraph*{Projection in $\{-\infty,0,1\}$}

We denote $\overline{-\infty}= -\infty$, $\overline{0}= 0$ and for any positive integer $x$, $\overline{x}= \overline{\infty} = 1$. For a matrix $M$ in $\matrices{i}{j}$, we denote by $\overline{M}$ the matrix $M$ where the coefficients are replaced by their barred version.

Let $\overline{\Omega}$ be the semiring $\left(\{-\infty,0,1\}, \max, +\right)$ where the operations are defined as follows: the $\max$ operation is given by the order $-\infty < 0 < 1 $ and the sum operation is commutative and given by $-\infty + x = -\infty$ for any element $x$, $0 + x = x$ for any $x \in \{0,1\}$ and $1 + 1 = 1$.

Let $\matricesOmbar{i}{j}$ be the set of matrices of size $i \times j$ over the semiring $\overline{\Omega}$. We use again $\otimes$ to denote the product of matrices induced by the operations in $\overline{\Omega}$.

Note that $x \mapsto \overline{x}$ is a morphism over $\Nmax$, as well as over $\matricesOm{i}{j}$. We will use the following basic properties without referencing them in the rest of the paper.

\begin{lem}\hfill
\label{lemma:propertyofbar}

\noindent Let $a,b\in \Nmax$, and let $\overline{a},\overline{b}$ be the projection of $a$ and $b$ into $\overline{\Omega}$, we have:
\begin{enumerate}[label={\normalfont(\arabic*)}]
\item If $a\le b$ then $\overline{a}\le \overline{b}$.
\item $\overline{a + b} = \overline{a} + \overline{b}$, where $\overline{a}+\overline{b}$ is taken  in the $\overline{\Omega}$ semiring.
\end{enumerate}
Let $M,N\in (\Nmax)^{d\times d}$ and let $\overline{M},\overline{N}$ be the pointwise projection of $M$ and $N$ into $\overline{\Omega}^{d\times d}$, we have:
\begin{enumerate}[resume,label={\normalfont(\arabic*)}]
\item $\overline{M\otimes N} = \overline{M}\otimes\overline{N}$ where $\overline{M}\otimes\overline{N}$ is taken  in the $\overline{\Omega}$ semiring.
\end{enumerate}
\end{lem}
\begin{proof}\hfill
\begin{enumerate}[label={\normalfont(\arabic*)}]
\item
\begin{itemize}
\item Suppose $a = -\infty$ then $-\infty  = a\le b$ and $-\infty=\overline{a}\le \overline{b}$.
\item Suppose $a= 0$, we have $a\le b \iff b\in\N$. Then $\overline{a} \le \overline{b}$. 
\item Suppose $a\in \N$, we have $a\le b \iff b\in \N_{\ge a}$, thus $\overline{b} = 1$. Then $1 = \overline{a} \le \overline{b}$.
\end{itemize}
\item
\begin{itemize}
\item  Suppose $a=-\infty$ (or, $b = -\infty$) then $a+b = \overline{a+b} = \overline{a} + \overline{b} = -\infty$.
\item Suppose $a=b=0$, then $a+ b= \overline{a+b} = \overline{a} + \overline{b} = 0$.
\item Suppose $a\in \mathbb{N}_{\ge 1}$, $b\in\N$. Then $a+b \in \mathbb{N}_{\ge 1}, \overline{a+b} = 1, \overline{a} = 1, \overline{a} + \overline{b} = 1$.
\end{itemize}
    \item We have \begin{align*}\qquad\overline{(M\otimes N)_{i,k}} &= \overline{M_{i,j} + N_{j,k}} \text{ for some } j
     \\&= \overline{M_{i,j}} + \overline{N_{j,k}} 
     \\&\le \max_{j}  \overline{M_{i,j}} + \overline{N_{j,k}}  
     \\&= (\overline{M}\otimes\overline{N})_{i,k}\end{align*}
    and
    \begin{align*}\qquad(\overline{M}\otimes\overline{N})_{i,k} &=  \overline{M_{i,j}} + \overline{N_{j,k}}  \text{ for some } j
    \\&= \overline{M_{i,j} + N_{j,k}}  
    \\&\le \overline{\max_{j} M_{i,j} + M_{j,k}} 
    \\&= \overline{(M\otimes N)_{i,k}}.\qedhere\end{align*}
\end{enumerate}
\end{proof}

Analogously to $(\ideal_{Q,i}, \otimes)$, given a finite set $Q$ and a positive integer $i$, we denote by $(\overline{\ideal_{Q,i}}, \otimes)$ the semigroup where:
\begin{itemize}
\item $\overline{\ideal_{Q,i}}$ is the union of the sets $Q \times \overline{\Omega} \times Q \times \matricesOmbar{i}{i}$ and $\{\bot\}$.
\item $(p,x,q,M)\otimes(p',x',q',M') = (p,x + x',q', M\otimes M')$ if $q = p'$ and $\bot$ otherwise.
\end{itemize}

\paragraph*{Semigroup of paths of $\autA$ and $\autB$}

Recall we have fixed a deterministic max-plus automaton $\autA = \langle Q_\autA, \Sigma, M_\autA, I_\autA, F_\autA\rangle$ and a max-plus automaton $\autB = \langle Q_\autB, \Sigma, M_\autB, I_\autB, F_\autB\rangle$ over the same alphabet $\Sigma$.

\begin{defi}\label{def:sgofp}
The \emph{semigroup of paths} of $\autA$ and $\autB$, denoted $\overline{\ideal}_{\autA,\autB}$, is the subsemigroup of $\overline{\ideal_{Q_\autA,|Q_\autB|}}$ generated by $\{(p,\overline{x},q,\overline{M_\autB(a)}) \mid a\in \Sigma, p\xrightarrow{a:x}q \text{ in } \autA\}$.
\end{defi}

\begin{runningexample}
From $a$ and $b$ respectively we construct
\[e_a = (p,1,p,
\begin{smallpmatrix}
0 & \nopath & \nopath & \nopath\\
\nopath & 1 & \nopath & \nopath\\
\nopath & \nopath & 0 & \nopath\\
\nopath & \nopath & \nopath & 0
\end{smallpmatrix}) \text{ and } e_b = (p,1,p,
\begin{smallpmatrix}
0 & 0 & \nopath & \nopath\\
\nopath & \nopath & 0 & \nopath\\
\nopath & \nopath & 0 & \nopath\\
\nopath & \nopath & \nopath & 1
\end{smallpmatrix}).\] 
Observe that $e_ae_a = e_a $, and, for example, $e_ae_b$ and $e_be_b$, in $\overline{\ideal}_{\autA,\autB}$, are given by  \[e_ae_b= (p,1,p,
\begin{smallpmatrix}
0 & 0 & \nopath & \nopath\\
\nopath & \nopath & 1 & \nopath\\
\nopath & \nopath & 0 & \nopath\\
\nopath & \nopath & \nopath & 1
\end{smallpmatrix}),  e_be_b=  (p,1,p,
\begin{smallpmatrix}
0 & 0 & 0 & \nopath\\
\nopath & \nopath & 0 & \nopath\\
\nopath & \nopath & 0 & \nopath\\
\nopath & \nopath & \nopath & 1
\end{smallpmatrix}).
\qedhere\] 
\end{runningexample}

\subsection{Semigroup of asymptotic behaviours}

The idempotent elements of a semigroup are elements $e$ such that $e\otimes e=e$. A matrix $M$ in $\matricesOm{i}{i}$ is called \emph{path-idempotent} if $\overline{M}$ is idempotent in $\matricesOmbar{i}{i}$. Similarly, an element $(p,x,q,M)$ of $\ideal_{Q,i}$ is called \emph{path-idempotent} if $p=q$ and $M$ is path-idempotent.

\paragraph*{Stabilisation operation}
The semigroup $\ideal_{Q,i}$ is equipped with a unary operation on its path-idempotent elements, called the \emph{stabilisation operation} and defined as follows: The stabilisation of elements in $\Omega$ is defined as: $(-\infty)^\sharp = -\infty$, $0^\sharp = 0$ and $1^\sharp = \infty^\sharp = \infty$. 
Given a path-idempotent matrix $M$ in $\matricesOm{i}{i}$, the stabilisation of $M$, denoted $M^\sharp$ is defined as the product $M\otimes M' \otimes M$ where $M'$ is the matrix $M$ where all the diagonal elements are replaced by their stabilisation.

\begin{defi}\label{def:op:stab}
The \emph{stabilisation operation} of $\ideal_{Q,i}$ is defined on its path-idempotents as follows: $(p,x,p,M)^\sharp = (p,x^\sharp,p,M^\sharp)$.
\end{defi}

\begin{runningexample}
Observe $e_a$ is idempotent and \[e_a^\#= (p,\infty,p,
\begin{smallpmatrix}
0 & \nopath & \nopath & \nopath\\
\nopath & \infty & \nopath & \nopath\\
\nopath & \nopath & 0 & \nopath\\
\nopath & \nopath & \nopath & 0
\end{smallpmatrix})\] indicating that the sequence of words $(a^n)_n$ has unbounded values in $\autA$, and the sequence $\left(M_{\autB}(a^n)_{q_2,q_2}\right)_n$ is also unbounded.

Consider \[e_a^\#e_b= (p,\infty,p,
\begin{smallpmatrix}
0 & 0 & \nopath & \nopath\\
\nopath & \nopath & \infty & \nopath\\
\nopath & \nopath & 0 & \nopath\\
\nopath & \nopath & \nopath & 1
\end{smallpmatrix}).\]
Recall from~\cref{re:notbig-O} we are aiming to represent a word of the shape $(a^nb)^n$. Hence, we would expect to iterate $e_a^\#e_b$ again, however it is not path-idempotent. Instead we can make one iteration manually resulting in $e_a^\#e_be_a^\#e_b$ which is idempotent,
\[e_a^\#e_be_a^\#e_b= (p,\infty,p,
\begin{smallpmatrix}
0 & 0 & \infty & \nopath\\
\nopath & \nopath & \infty & \nopath\\
\nopath & \nopath & 0 & \nopath\\
\nopath & \nopath & \nopath & 1
\end{smallpmatrix}).\]
Let us consider the effect of stabilisation on it. We have \[(e_a^\#e_be_a^\#e_b)^\#= (p,\infty,p,
\begin{smallpmatrix}
0 & 0 & \infty & \nopath \\
\nopath & \nopath & \infty & \nopath\\
\nopath & \nopath & 0 & \nopath\\
\nopath & \nopath & \nopath & \infty
\end{smallpmatrix}).\] 
The purpose of this operation is to identify unbounded entries after arbitrarily many iterations of either the first, the second, or both, stabilisation operations. Note that $\funcA((a^{n}ba^{n}b)^n) = 2n^2+2n$ while $M_{\autB}((a^{n}ba^{n}b)^n)_{q_1,q_3} =M_{\autB}((a^{n}ba^{n}b)^n)_{q_2,q_3}  =n$ and $M_{\autB}((a^{n}ba^{n}b)^n)_{q_4,q_4}= 2n$. Both sequences are unbounded as witnessed by the $\infty$ but this does not allow us to identify the different rates of growth between the entries. 
\end{runningexample}

\paragraph*{Flattening operation}
The semigroup $\ideal_{Q,i}$ is also equipped with another unary operation on its path-idempotent elements, called the flattening operation and defined as follows: Given a path-idempotent matrix $M$ in $\matricesOm{i}{i}$, the \emph{flattening} of $M$, denoted $M^\flat$ is defined as the product $\overline{M}\otimes \langle M^3 \rangle \otimes \overline{M}$ where $\langle M \rangle$ is the matrix $M$ where all the non diagonal elements are replaced by their barred version.

\begin{defi}\label{def:op:flat}
The \emph{flattening operation} of $\ideal_{Q,i}$ is defined on its path-idempotents as follows: $(p,x,p,M)^\flat = (p,x,p,M^\flat)$.
\end{defi}

\begin{runningexample}
Let us consider the effect of flattening on $e_a^\#e_be_a^\#e_b$:
 \[
(e_a^\#e_be_a^\#e_b)^\flat= (p,\infty,p,
\begin{smallpmatrix}
0 & 0 & 1 & \nopath\\
\nopath & \nopath & 1 & \nopath\\
\nopath & \nopath & 0 & \nopath\\
\nopath & \nopath & \nopath & 1
\end{smallpmatrix}).
\]
Here observe that $\infty$ corresponds to an entry with growth rates $n^2$ on the sequence of words $(a^nba^nb)^n$. Other entries, even unbounded sequences, with asymptotically smaller growth rates -- such as $n$ -- are projected to $1$ (or $0$). The flattening behaviour allows us to capture differences in growth rates (not fully, but enough for our purpose), by, roughly speaking, keeping only the fastest growing elements. Intuitively $(e_a^\#e_be_a^\#e_b)^\flat$ demonstrates that $\autA$ can grow faster than $\autB$ and so $\autA$ is not big-O of $\autB$, we formalise this intuition as a witness of non-domination in the next section.
\end{runningexample}

\paragraph*{The semigroup of asymptotic behaviours of $\autA$ and $\autB$}

Recall we have fixed a deterministic max-plus automaton $\autA = \langle Q_\autA, \Sigma, M_\autA, I_\autA, F_\autA\rangle$ and a max-plus automaton $\autB = \langle Q_\autB, \Sigma, M_\autB, I_\autB, F_\autB\rangle$ over the same alphabet $\Sigma$.

\begin{defi}\label{def:sgab}
The \emph{semigroup of asymptotic behaviours} of $\autA$ and $\autB$, denoted $\ideal_{\autA,\autB}$, is the subsemigroup of $\ideal_{Q_\autA,|Q_\autB|}$ generated by $\{(p,\overline{x},q,\overline{M_\autB(a)}) \mid a\in \Sigma, p\xrightarrow{a:x}q \text{ in } \autA\}$ and closed under the stabilisation and flattening operations.
\end{defi}

\section{Decision Procedure}
\label{sec:decision}

\subsection{Witnesses}

Given a deterministic max-plus automaton $\autA$ and a max-plus automaton $\autB$ over the same alphabet, an element $(p,x,q,M)$ in $\ideal_{\autA,\autB}$ is called a \emph{witness of non-domination} if:
\begin{itemize}
\item $p$ is initial in $\autA$,
\item $q$ is final in $\autA$,
\item $x = \infty$,
\item $\overline{I_\autB} \otimes M \otimes \overline{F_\autB} < \infty$.
\end{itemize}

\begin{thm}\label{thm:2assertion}
Given two max-plus automata $\autA,\autB$ such that $\autA$ is deterministic and $\funcB:\Sigma^* \to \N$, the two following assertions are equivalent:
\begin{itemize}
\item $\autA$ is big-O of $\autB$.
\item There is no witness of non-domination in $\ideal_{\autA,\autB}$.
\end{itemize}
\end{thm}
We will not prove \cref{thm:2assertion} directly, rather we will prove a refined version of the theorem incorporating tractable witnesses, which are defined next.

\subsection{Tractable witnesses}

A witness could be any element in $\ideal_{\autA,\autB}$, found through arbitrary application of product, stabilisation and flattening, satisfying the conditions. We now consider a restricted form of witness in which we place a restriction on the sequence of operations to construct it. 
\begin{defi}[Tractable witness of non-domination]\label{def:tractable_witness}
We say an element $g$ in $\ideal_{\autA,\autB}$ is a \emph{tractable witness of non-domination} if it is both a witness of non-domination and of the form \[g = g_0 (g_1 (\dots (g_{k-2} ( g_{k-1} (g_{k})^\#g'_{k-1})^\flat g'_{k-2})^\flat \dots)^\flat g'_{1})^\flat g'_{0},\] for $k \le 3|\overline{\ideal}_{\autA,\autB}|$ and $g_i, g'_i \in \overline{\ideal}_{\autA,\autB}\cup\{\monoidid\}$, where $\monoidid$ is an added identity element of $\ideal_{\autA,\autB}$ such that $e\otimes \monoidid = \monoidid \otimes e = e$ for all $e\in \ideal_{\autA,\autB}$.
\end{defi}
\begin{runningexample}\label{re:actualtractablewitness}
$(e_a^\#e_be_a^\#e_b)^\flat$ is a witness but not a tractable witness of non-domination. However, $(e_ae_be_a^\#e_b)^\flat$ will turn out to be a tractable witness. Intuitively, it represents the sequence of words $(aba^nb)^n$, which is almost the same as the sequence used to show $\autA$ is not big-O of $\autB$ in \cref{re:notbig-O}.
\end{runningexample}

We will now strengthen \cref{thm:2assertion}, in which we add the condition that there is a tractable witness;  this allows us to limit our search to tractable witnesses.
\begin{thm}\label{thm:3assertion}
Given two max-plus automata $\autA,\autB$ such that $\autA$ is deterministic and $\funcB:\Sigma^* \to \N$, the following assertions are equivalent:
\begin{itemize}
\item $\autA$ is big-O of $\autB$.
\item There is no witness of non-domination in $\ideal_{\autA,\autB}$.
\item There is no tractable witness of non-domination in $\ideal_{\autA,\autB}$.
\end{itemize}
\end{thm}
The benefit of a tractable witness will be that we can identify the existence of one in PSPACE. In the next section we present the PSPACE algorithm to detect a tractable witness and then we will prove the equivalences of \cref{thm:3assertion} to conclude the result.

\subsection{PSPACE algorithm}

We define a non-deterministic procedure to construct a tractable witness from middle out, that runs in polynomial space. Since NPSPACE = PSPACE (from Savitch's theorem), this will allow us to conclude.

Any element $g \in \overline{\ideal}_{\autA,\autB}$ can be constructed using at most ${|\overline{\ideal}_{\autA,\autB}|}$ product operations from the generators. Suppose $g= g_1\otimes \dots\otimes g_m$ such that $g_i$ are generators and $m$ is minimal, then we can assume that $g_1\otimes \dots\otimes g_i$ is different from  $g_1\otimes \dots\otimes g_j$ for each $i\ne j, i,j\le m$. Thus $m \le {|\overline{\ideal}_{\autA,\autB}|}$.

The procedure is as follows:
\begin{itemize} 
    \item Non-deterministically choose $k \le 3|\overline{\ideal}_{\autA,\autB}|$.
    \item Let $g$ be a non-deterministically chosen  idempotent element of $\overline{\ideal}_{\autA,\autB}$, constructed in at most ${|\overline{\ideal}_{\autA,\autB}|}$ steps.
    \item Update $g$ to be the stabilisation $g^\#$.
    \item Repeating for $i = k$ to $ i = 0$, we update $g$ with $(g_i \  g \ g'_{i})^\flat$ for some $g_i,g_i'\in  \overline{\ideal}_{\autA,\autB}\cup\{\monoidid\}$ in the following way:
    \begin{itemize}
    \item Update $g$ by non-deterministically choosing a generator or $\monoidid$ and multiply on the left of $g$. Repeat up to ${|\overline{\ideal}_{\autA,\autB}|}$ many times.
    \item Update $g$ by non-deterministically choosing a generator or $\monoidid$ and multiply on the right of $g$. Repeat up to ${|\overline{\ideal}_{\autA,\autB}|}$ many times.
    \item Except for $i = 0$, check that $g$ is path-idempotent and update $g$ to be the flattening of \nolinebreak $g$.
    \end{itemize}
    \item Check if $g$ is a witness.
\end{itemize}
At any moment we are only storing one element $g$ of $\ideal_{\autA,\autB}$, plus the space needed for doing the product, iteration and stabilisation operations, the  current iteration and number of iterations $i$ and $k$, and how many elements we have multiplied (on the left or on the right) with $g$. This requires only polynomial space.
This results in an NPSPACE algorithm, which is equivalent to a PSPACE algorithm.

\subsection{Proof of Theorem~\ref{thm:3assertion}}

To prove \cref{thm:3assertion}, we are going to prove the following result - the notions of factorisation trees and faults will be introduced in due course.

\begin{thm}\label{thm:4assertion}
Given two max-plus automata $\autA,\autB$ such that $\autA$ is deterministic and $\funcB:\Sigma^* \to \N$, the following assertions are equivalent:
\begin{enumerate}[label={\normalfont(\arabic*)}]
\item $\autA$ is not big-O of $\autB$.
\item There is a witness of non-domination in $\ideal_{\autA,\autB}$.
\item There is a tractable witness of non-domination in $\ideal_{\autA,\autB}$.
\item Some word has a factorisation tree, of height at most $3|\overline{\ideal}_{\autA,\autB}|$, with a fault.
\end{enumerate}
\end{thm}

Since a tractable witness is a special case of a witness, it is clear that (3) implies (2). The remainder of the paper will prove the remaining implications. 

In \cref{section:facto-fault}, we introduce the notions of factorisation trees and faults. In \cref{1implies4}, we prove that (1) implies (4) - by proving its contrapositive. In \cref{4implies3}, we prove that (4) implies (3). \cref{1implies4,4implies3} are independent of each other, but rely on \cref{section:facto-fault}. Finally, in \cref{2implies1}, we prove that (2) implies (1). This later section can be read independently of the other ones.

\section{Factorisation trees and faults}
\label{section:facto-fault}

Recall we have fixed a deterministic max-plus automaton $\autA = \langle Q_\autA, \Sigma, M_\autA, I_\autA, F_\autA\rangle$ and a max-plus automaton $\autB = \langle Q_\autB, \Sigma, M_\autB, I_\autB, F_\autB\rangle$ over the same alphabet $\Sigma$.

\subsection{Factorisation trees}
Let $w = w_1\cdots w_k$ with $w_1, \ldots, w_k \in \Sigma$ such that $\funcA(w) \neq -\infty$ and let:
$$p_0 \xrightarrow{w_1:x_1} p_1 \xrightarrow{w_2:x_2} p_2 \dotsm p_{k-1}\xrightarrow{w_{k}:x_k} p_k$$ 
be its unique accepting path in $\autA$. Let $\alpha_w(w_i)$ be the element of $\overline{\ideal}_{\autA,\autB}$ defined as $(p_{i-1},\overline{x_i},p_i,\overline{M_\autB(w_i)})$.

A factorisation tree on $w$ is a finite ordered tree in which every node $\nu$ in the tree is labelled by an element in $\overline{\ideal}_{\autA,\autB}$, denoted $\alpha(\nu)$, such that:
\begin{itemize}
\item there are $k$ leaves labelled with $\alpha_w(w_1), \ldots, \alpha_w(w_k)$, 
\item internal nodes have two or more children, and are labelled by the product of the labels of their children: a node $\nu$ with children $\nu_1,\dots \nu_m$ for some $m\ge 2$, is labelled with $\alpha(\nu) = \alpha(\nu_1)\otimes \alpha(\nu_2) \otimes \dots \otimes \alpha(\nu_m)$. A node with two children is called a product node,
\item if a node has at least three children then the children and the node are all labelled by the same idempotent element - such a node is called an idempotent node (in particular, $\alpha(\nu) = \alpha(\nu_i)$ for all $i\le m$).
\end{itemize}

Note that for a word $w$ such that $\funcA(w) \neq -\infty$, no node in a factorisation tree on $w$ can be labelled by $\bot$. It is also clear that the labelling of the root of the subtree with leaves $\alpha_w(w_i), \ldots, \alpha_w(w_j)$ for some $i<j$, corresponds to the element in the semigroup of paths witnessing the existence of $0$ or positive weights paths in $\autA$ and $\autB$ on the word $w_i \cdots w_j$.

\begin{thm}[Simon's Factorisation Theorem~\cite{Simon90}]
\label{theorem:Simon90}
There exists a positive integer $H$ such that for all $w\in\Sigma^*$, there exist a factorisation tree on $w$ of height at most $H$.
\end{thm}

Note that $H$ does not depend on the word $w$, only on the size of $\overline{\ideal}_{\autA,\autB}$. Further we have that $H \le 3|\overline{\ideal}_{\autA,\autB}|-1$ due to the bound of \cite{Kufleitner08}, which is tighter than the $3|\overline{\ideal}_{\autA,\autB}|$ bound of Colcombet~\cite{Colcombet07}, and the original bound of $9|\overline{\ideal}_{\autA,\autB}|$  by Simon.

\subsection{Contributors}

Let $t$ be a factorisation tree on a word $w$, such that $\funcA(w) \neq -\infty$. For each node $\nu$ of the tree, we define its set of contributors $C_{\nu}$ as follows, in a top-down manner:
\begin{itemize}
\item if the root is labelled $(p,x,q,M)$, the contributors of the root is the set of pairs $(i,j)$ such that $i$ is initial in $\autB$, $j$ is final in $\autB$ and $M_{i,j} \neq - \infty$.
\item if a node has a set of contributors $C$, and has two children labelled $(p,x,q,M)$ and $(q,x,r,P)$,
\begin{itemize}
\item  the set of contributors of the left child is: \\
$\{(i,\ell)\mid \exists j : (i,j) \in C, P_{\ell,j} \neq -\infty, M_{i,\ell}\ne-\infty\}$,
\item  the set of contributors of the right child is:\\
$\{(\ell,j)\mid \exists i : (i,j) \in C, M_{i,\ell} \neq -\infty, P_{\ell,j}\ne-\infty\}$.
\end{itemize}
\item if a node has a set of contributors $C$ and has at least three children, labelled by an idempotent element $(p,x,p,M)$, then:
\begin{itemize}
\item the left-most child has set of contributors:\\ $\{(i,\ell)\mid\exists j: (i,j) \in C, M_{\ell,j} \neq -\infty , M_{i,\ell}\ne-\infty\}$,
\item the right-most child has set of contributors:\\ $\{(\ell,j)\mid \exists i : (i,j) \in C, M_{i,\ell} \neq -\infty, M_{\ell,j}\ne-\infty\}$,
\item the other children have set of contributors:\\
 $\{(\ell,k)\mid  \exists (i,j) \in C, M_{i,\ell} \neq -\infty, M_{k,j} \neq -\infty, \\ M_{\ell,k}\ne-\infty,M_{k,\ell}\ne-\infty\}$. 
\end{itemize}
\end{itemize}

Contributors indicate which elements of the matrix meaningfully contribute to a valid run for the whole word. For example, at the root, not every entry is a weight on a run from an initial state to a final state, so not all entries contribute to the value computed on the word. The choices for the root and product nodes are uncontroversial, however the choice is non-trivial for the middle children of an idempotent node. Here only entries that could be repeated many times are taken; this is because, whilst other transitions could contribute to a valid run, they could contribute only once, whereas the entries we consider can be used in many of the idempotent children.  

\begin{figure}
         \centering
         \includegraphics[width=0.7\linewidth]{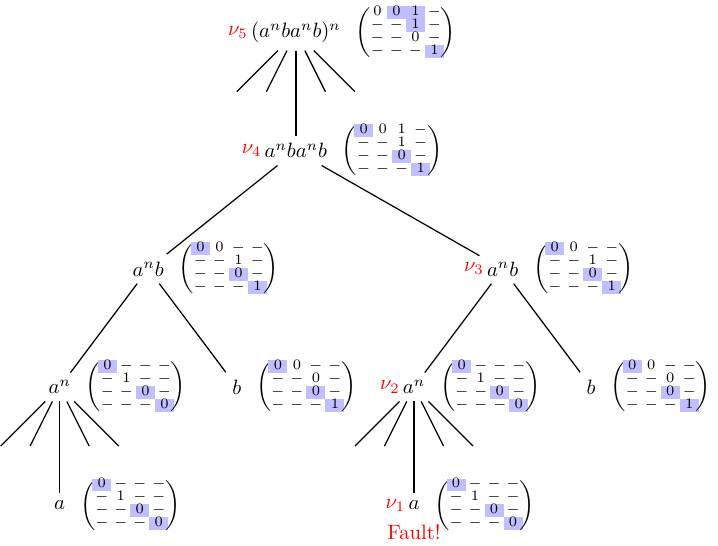}
         \caption{A possible factorisation tree for the word $w= (a^nba^nb)^n$. Nodes are labelled by the sub-word of $w$  and the path-behaviour of $M_{\autB}$. The contributors of each node are indicated by highlighting the corresponding matrix entries. A node with a fault is also indicated, inducing the sequence of nodes $\nu_1,\dots, \nu_5$ from the fault to the root.}
         \label{fig:tree}
\end{figure}

\begin{runningexample}

\cref{fig:tree} depicts a factorisation tree for the word $(a^nba^nb)^n$ of height 5. Every node is depicted with the sub-word and a partial description of its $\alpha$-labelling: every node is $\alpha$-labelled by $(p,1,p,M)$, where $M$ is the matrix depicted.
The contributors are highlighted in the matrix for each node.  Only a representation of the middle children is depicted for idempotent nodes. We will return to the fault label and the sequence $\nu_1,\dots,\nu_5$ after defining faults.
\end{runningexample}

We will make use of the property that every node has at least one contributor, stated in the following proposition.
\begin{prop}
\label{prop:nodeemptyparentempty}
Given a word $w$ such that $\funcB(w) \neq -\infty$, every node in any factorisation tree on $w$ has a non non-empty of contributors.
\end{prop}

\begin{proof}
We show that if a node has a contributor then all its children have a contributor. Since we assume $I_\autB \otimes M_\autB(w)\otimes F_\autB\ne-\infty$ we observe that the root in any factorisation tree has non-empty contributors, this would imply that every element of the tree has non-empty contributors.

First observe, that the definition of a contributor requires that for every $\nu$, with $\alpha(\nu) = (p,x,q,M)$, if $(i,j)\in C_\nu$ then $M_{i,j}\ne -\infty$. We consider two cases for both type of internal node. Suppose $(i,j) \in C_{\nu}$ and $M_{i,j}\ne-\infty$. 

\begin{case}[$\nu$ is a product node]Suppose  $\alpha(\nu) = (p,x,q,M)$ is the product of two children  $\nu_1$ and $\nu_2$ with $\alpha(\nu_1) = (p,x_1,r,P_1)$ and $\alpha(\nu_2) = (r,x_2,q,P_2)$ then $M = P_1\otimes P_2$, and $M_{i,j} = (P_1)_{i,k}+ (P_2)_{k,j}$ for some $k$. Thus $(i,k)\in C_{\nu_1}$ and $(k,j) \in C_{\nu_2}$.
\end{case}
\begin{case}[$\nu$ is an idempotent node] Suppose $\nu$ has children $\nu_1,\dots, \nu_k$ for $k\ge 3$, with $\alpha(\nu) = \alpha(\nu_i) = (p,x,p,M)$ for all $1\le i\le k$.
Since $M$ is idempotent, we have $M = M^d$ for any choice of $d\ge 1$. Note that there is no requirement that $d\le k$, the equivalence holds for \emph{all} $d$ by idempotence.
In particular, let us fix a choice of some $d\ge |Q_\autB| + 3$.
Choose a contributor of $\nu$, $(i,j) \in C_{\nu}$, and so we have $M_{i,j} \ne -\infty$. Further, since $M$ is idempotent, we have $M_{i,j} = M^{d}_{i,j}$. Thus there exists a sequence $i = \ell_1,\dots,\ell_{d+1}=j$ such that $M^{d}_{i,j} = M_{\ell_1,\ell_2} + M_{\ell_2,\ell_3} + \dots + M_{\ell_{d},\ell_{d+1}}$, where $M_{\ell_n,\ell_{n+1}} \ne -\infty$ for every $n\le d$.  

 Note then that $(\ell_1,\ell_2) \in C_{\nu_1}, (\ell_{d},\ell_{d+1}) \in C_{\nu_k}$.
 Furthermore, by simple application of the pigeon hole principle, there exists \emph{distinct} indices $n,m$ such that $i_n = i_m$ (as $d \ge |Q_\autB|+3$). Thus we have $(i_n,i_n) \in C_{\nu_2}\cap \dots\cap C_{\nu_{k-1}}$.\qedhere
\end{case}
\end{proof}

\subsection{Faults}

Given a word $w$ such that $\funcA(w) \neq -\infty$ and a factorisation tree on $w$, a node labelled with $(p,x,q,M)$, is called a \emph{fault} if:
\begin{itemize}
\item it is the child of an idempotent node, but is neither the left-most nor the right-most child,
\item $x = 1$,
\item $M_{i,j} = 0$ for all pairs $(i,j)$ in its set of contributors.
\end{itemize}

\begin{runningexample}
Let us return to the factorisation tree for $(a^nba^nb)^n$ depicted in \cref{fig:tree}. Observe that the node indicated by $\nu_1$ is a middle-child of an idempotent with only zero entries in the contributors and is therefore a fault.
Since the two subtrees below $\nu_4$ are identical the other node labelled by $a$ is also a fault. In \cref{runningeg:constructingawitnessfromfault}  we will use the  indicated fault $\nu_1$ to construct a (tractable) witness of non-domination.
\end{runningexample}

\section{No tree has a fault implies Big-O}
\label{1implies4}
\setcounter{case}{0}

In this section, we suppose that no word has a factorisation tree of height at most $3|\overline{\ideal}_{\autA,\autB}|$ with a fault, and we construct a positive integer $c$ such that: 
\begin{equation}\label{eq:bigoformula}
\funcA(w) \leq c\funcB(w) + c \quad \text{ for all } w\in\Sigma^*.
\end{equation}

Recall that we can also assume that $\funcB(w) \neq -\infty$ for all words $w$.

Let $w = w_1\cdots w_k$ with $w_1, \ldots, w_k \in \Sigma$ such that $\funcA(w) \neq -\infty$ and let:
$$p_0 \xrightarrow{w_1:x_1} p_1 \xrightarrow{w_2:x_2} p_2 \dotsm p_{k-1}\xrightarrow{w_{k}:x_k} p_k$$ 
be its unique accepting path in $\autA$. Given a factorisation tree $t$ on $w$, for a node $\nu$ in $t$, root of the subtree with leaves $\alpha_w(w_i), \ldots, \alpha_w(w_j)$, we denote by:
\begin{itemize}
\item $\aval(\nu)$ the weight $x_i+ \ldots + x_j$ of the path in $\autA$ corresponding to the factor $w_i \cdots w_j$, 
\item $\bval(\nu)$ the matrix $M_\autB(w_i \cdots w_j)$.
\end{itemize}

Let $\Lambda$ be the largest value occurring on a transition in $\autA$, and let $c_h = (4|Q_\mathcal{B}|+4)^{h}\Lambda$ for positive integers $h$. We prove the following property:

\begin{prop}
\label{prop:nofault}
Let $w$ be a word such that $\funcA(w) \neq -\infty$ and $t$ a factorisation tree on $w$ with no fault. Let $\nu$ be a node in $t$ of height $h$ for some positive integer $h$. Then:
\[ \aval(\nu) \leq c_h \max_{(i,j) \in C_{\nu}} \bval(\nu)_{i,j} + c_h\]
\end{prop}

Observe that \cref{eq:bigoformula} is trivial for $w$ such that $\funcA(w) = -\infty$. For $w$ such that $\funcA(w) \neq -\infty$,  \cref{eq:bigoformula} is a direct corollary of \cref{prop:nofault} choosing $c=c_H$ where $H = 3|\overline{\ideal}_{\autA,\autB}|$, $\nu$ as the root of a factorisation tree on $w$ of height at most $H$, which exists by \cref{theorem:Simon90}.

\begin{proof}
The proof is by induction on $h$.
\begin{case}[If $\nu$ is a leaf and $h=0$]
By definition of $\Lambda$ as the largest value occurring on a transition in $\autA$, and by definition of $c_0$, we have:
 \[\aval(\nu) \le \Lambda \le c_0 \le c_0 \max_{(i,j) \in C_\nu} \bval(\nu)_{i,j} + c_0.\] 
The last inequality holds only if the set of contributors $C_\nu$ is not empty, which is the case by \cref{prop:nodeemptyparentempty} since we assume $\funcB(w) \ge 0$ for all $w\in \Sigma^*$.
\end{case}

\begin{case}[If $\nu$ is a product node]
Let $\nu_1$ and $\nu_2$ be the two children of $\nu$. Then $\aval(\nu) = \aval(\nu_1) + \aval(\nu_2)$. By induction, for each child $m\in\{1,2\}$ we have:
\[ \aval(\nu_m) \leq c_{h-1} \max_{(i,j) \in C_{\nu_m}} \bval(\nu_m)_{i,j} + c_{h-1}\]
Suppose $\aval(\nu_1) \geq  \aval(\nu_2)$ (the case $\aval(\nu_2) >  \aval(\nu_1)$ is symmetric). Then:
\begin{align*}
\aval(\nu_1)+\aval(\nu_2) & \le 2\aval(\nu_1) \\
& \le 2c_{h-1} \max_{(i,j) \in C_{\nu_1}} \bval(\nu_1)_{i,j} + 2c_{h-1}
\end{align*}
Consider $(d,f) \in C_{\nu_1}$ for which this maximum is attained. Since $(d,f)$ is a contributor of a left child then there exists $g$ such that $\bval(\nu_2)_{f,g} \neq -\infty$ and $(d,g)$ is a contributor of $\nu$.
We get:
\begin{align*}
\aval(\nu) &\le \aval(\nu_1) + \aval(\nu_2)
\\&\le 2c_{h-1} \bval(\nu_1)_{d,f} + 2c_{h-1}
\\&\le 2c_{h-1} (\bval(\nu_1)_{d,f} + \bval(\nu_2)_{f,g}) + 2c_{h-1}
\\&\le 2c_{h-1}  \max_{(i,j) \in C_{\nu}} \bval(\nu)_{i,j} + 2c_{h-1}
\\&\le c_h  \max_{(i,j) \in C_{\nu}} \bval(\nu)_{i,j} + c_{h} \tag{as $c_h > 2c_{h-1}$.}
\end{align*}
\end{case}

\begin{case}[If $\nu$ is an idempotent node]
Let $\nu_1,\dots,\nu_d$ be the children of $\nu$. 
By definition and idempotency, $\overline{\aval(\nu)} = \overline{\aval(\nu_m)}$ for all $m=1,\ldots,d$. If $\overline{\aval(\nu)} = 0$, then we directly get the result since the set of contributors $C_\nu$ is not empty by \cref{prop:nodeemptyparentempty} (because we assume that $\funcB(w) \ge 0$ for all $w\in \Sigma^*$). Let's suppose now that $\overline{\aval(\nu)} = \overline{\aval(\nu_m)} = 1$ for all $m=1,\ldots,d$.

By inductive hypothesis, for all $m=1,\ldots, d$, we have:
\[ \aval(\nu_m) \leq c_{h-1} \max_{(i,j) \in C_{\nu_m}} \bval(\nu_m)_{i,j} + c_{h-1}.\]
Since there is no fault in the tree, then for all $m$, there is some $(i,j) \in C_{\nu_m}$ such that  $\bval(\nu_m)_{i,j} \geq 1$, and hence:
\begin{equation}\label{eq:subchildrennoplusc} \aval(\nu_m) \leq 2c_{h-1} \max_{(i,j) \in C_{\nu_m}} \bval(\nu_m)_{i,j}.\end{equation}

Let $(i_m,j_m)$ be a pair in $C_{\nu_m}$ on which this maximum is attained. Note that $\overline{\bval(\nu_m)}$ are the same for all $m$ as $\overline{\bval(\nu)}$, and that this is an idempotent matrix. Let us define $i\sim j$ if and only if $\overline{\bval(\nu)_{i,j}} = \overline{\bval(\nu)_{j,i}} \neq -\infty$. By idempotency, this gives an equivalence relation, and we denote by $\mathcal{S}_1,\dots,\mathcal{S}_{z}$ its equivalence classes. Note that $z$ is bounded by the number of states of $\autB$.
We now partition the set $\{1,\ldots, d\}$ (the children of the node $\nu$) as the union $\Phi$ of the sets $\{1\}$, $\{d\}$, $\Gamma_{\mathcal{S}_f,\text{even}}$ and $\Gamma_{\mathcal{S}_f,\text{odd}}$ for all $f=1,\ldots,z$, where:
\[\Gamma_{\mathcal{S}_f,\text{even}} =  \{ m \in\{2,\dots, d-1\} \mid i_m,j_m\in \mathcal{S}_f \text{ and $m$ is even}\}\]
and
\[\Gamma_{\mathcal{S}_f,\text{odd}} =  \{ m \in\{2,\dots, d-1\} \mid i_m,j_m\in \mathcal{S}_f \text{ and $m$ is odd}\}.\]
This gives a partition of $\{1,\ldots, d\}$ since $(i_m,j_m)$ is in the set of contributors of $\nu_m$ and then $i_m$ and $j_m$ are in the same $\mathcal{S}_f$ for some $f$. We partition this way into even and odd indices to be able to reconstruct a path: if we select for example the even indices in one of the $\mathcal{S}_f$, by definition, we can construct a path in the automaton taking the transitions corresponding to these indices.
Note that the number of sets forming $\Phi$ is bounded by $2(|Q_\autB| + 1)$, where we recall that $|Q_\autB|$ is the number of states of $\autB$.

For each $\Gamma$ in $\Phi$, let $x_\Gamma = \sum_{m\in \Gamma} \aval(\nu_m)$, and denote by $\Gamma_{max}$ one for which this sum is maximal. Observe that:
\[\aval(\nu) \leq 2(|Q_\autB| + 1)x_{\Gamma_{max}}.\]

\begin{subcase}[$\Gamma_{max} = \Gamma_{\mathcal{S}_f,\text{odd}}$ for some $f$]
We have:
\begin{align*}
\aval(\nu) &\le 2(|Q_\autB| + 1)x_{\Gamma_{max}}
\\&= 2(|Q_\autB| + 1)\sum_{m\in \Gamma_{\mathcal{S}_f,\text{odd}}}\aval(\nu_m) 
\\&\le 2(|Q_\autB| + 1)\sum_{m\in \Gamma_{\mathcal{S}_f,\text{odd}}}(2c_{h-1} \bval(\nu_m)_{i_m,j_m})
\tag{by~\cref{eq:subchildrennoplusc}}
\\&= 4c_{h-1}(|Q_\autB| + 1)\sum_{m\in \Gamma_{\mathcal{S}_f,\text{odd}}}\bval(\nu_m)_{i_m,j_m}.
\end{align*}

By definition of contributors, and construction of $\Gamma_{\mathcal{S}_f,\text{odd}}$, there exists $(i,j)$ in $C_{\nu}$ and $i=\ell_0, \ell_1, \ldots, \ell_d=j$ such that $i_m = \ell_{m-1}$ and $j_m = \ell_{m}$ for all $m$ in $\Gamma_{\mathcal{S}_f,\text{odd}}$ and $\bval(\nu_m)_{\ell_{m-1},\ell_{m}} \neq -\infty$ for all $m=1,\ldots,d$, hence 
\[\sum_{m\in \Gamma_{\mathcal{S}_f,\text{odd}}}\bval(\nu_m)_{i_m,j_m} \leq \sum_{m=1}^{d}\bval(\nu_m)_{\ell_{m-1},\ell_{m}}.\]
Since we also have:
$$\max_{(i,j) \in C_{\nu}} \bval(\nu)_{i,j} = \max_{(i,j) \in C_{\nu}}\  \max_{i=\ell_0, \ell_1, \ldots, \ell_d=j}\left(\sum_{m=1}^{d}\bval(\nu_m)_{\ell_{m-1},\ell_{m}}\right)$$
we obtain:
\begin{align*}
\aval(\nu) &\le 4c_{h-1}(|Q_\autB| + 1)\sum_{m\in \Gamma_{\mathcal{S}_f,\text{odd}}}\bval(\nu_m)_{i_m,j_m}
\\&\le 4c_{h-1}(|Q_\autB| + 1)\max_{(i,j) \in C_{\nu}} \bval(\nu)_{i,j}.
\end{align*}
\end{subcase}

\begin{subcase}[$\Gamma_{max} = \Gamma_{\mathcal{S}_f,\text{even}}$ for some $f$]
This is similar to the previous case.
\end{subcase}

\begin{subcase}[$\Gamma_{max} = \{1\}$]
This is similar to the product case. We have:
\begin{align*}
\aval(\nu) &\le 2(|Q_\autB| + 1)x_{\Gamma_{max}}
\\&= 2(|Q_\autB| + 1)\aval(\nu_1)
\\&\le 4(|Q_\autB| + 1)c_{h-1} \max_{(i,j) \in C_{\nu_1}} \bval(\nu_1)_{i,j}
\\&\le 4(|Q_\autB| + 1)c_{h-1} \max_{(i,\ell) \in C_{\nu}} \bval(\nu)_{i,\ell}  \tag{by definition of contributors of the left-most child.}
\end{align*}
\end{subcase}

\begin{subcase}[$\Gamma_{max} = \{d\}$]
This is symmetric to the previous case.\qedhere
\end{subcase}
\end{case}
\end{proof}

\section{Construction of a witness if there is a tree with a fault}
\label{4implies3}

We suppose that some word has a factorisation tree of height at most $3|\overline{\ideal}_{\autA,\autB}|$ with a fault. We are going to construct a tractable witness of non-domination in $\ideal_{\autA,\autB}$.

Let $\nu$ be a fault of maximal height in this tree. Let $\nu_1 = \nu$ and let $\nu_{h+1}$ be the direct parent of $\nu_h$ for $h=2,\dots,m$, where $\nu_m$ is the root node. Before giving the construction of the tractable witness formally, we will see how to construct it on our running example.

\begin{runningexample}\label{runningeg:constructingawitnessfromfault}

We consider the tree depicted in \cref{fig:tree}. We define $\nu_1, \ldots, \nu_5$ as explained above. A tractable witness is constructed by doing the stabilisation operation on the labelling of $\nu_1$, then doing the product on the right with the labelling corresponding to $b$ and on the left with the one corresponding to $ab$. This gets us to $\nu_4$. At this point, we take the flattening of what we have obtained. We define the $\beta$-labelling of the nodes $\nu_h$:

\[
\beta(\nu_2) =(p,1, p,  \begin{smallpmatrix}
  0 & \nopath & \nopath & \nopath\\
  \nopath & 1 & \nopath & \nopath\\
  \nopath & \nopath & 0 & \nopath\\
  \nopath & \nopath & \nopath & 0
  \end{smallpmatrix}  )^\sharp 
= (p,\infty, p, \begin{smallpmatrix}
  0 & \nopath & \nopath & \nopath\\
  \nopath & \infty & \nopath & \nopath\\
  \nopath & \nopath & 0 & \nopath\\
  \nopath & \nopath & \nopath & 0
  \end{smallpmatrix})
\]
\[
\begin{split} 
\beta(\nu_3) = & \beta(\nu_2) \otimes (p,1, p, \begin{smallpmatrix}
  0 & 0 & \nopath & \nopath\\
  \nopath & \nopath & 0 & \nopath\\
  \nopath & \nopath & 0 & \nopath\\
  \nopath & \nopath & \nopath & 1
  \end{smallpmatrix}) \\
= & (p,\infty, p, \begin{smallpmatrix}
  0 & 0 & \nopath & \nopath\\
  \nopath  & \nopath & \infty & \nopath\\
  \nopath & \nopath & 0 & \nopath\\
  \nopath & \nopath & \nopath & 1
  \end{smallpmatrix})
\end{split}
\]
\[
\begin{split} 
\beta(\nu_4) = &
(p,1,p,\begin{smallpmatrix}
   0 & 0 & \nopath & \nopath\\
   \nopath  & \nopath & 1 & \nopath\\
   \nopath & \nopath & 0 & \nopath\\
   \nopath & \nopath & \nopath & 1
   \end{smallpmatrix} ) \otimes \beta(\nu_3) \\
= &
 (p,\infty, p, \begin{smallpmatrix}
0 & 0 & \infty & \nopath\\
  \nopath  & \nopath & \infty & \nopath\\
  \nopath & \nopath & 0 & \nopath\\
  \nopath & \nopath & \nopath & 1
  \end{smallpmatrix})
\end{split}
\]
\[
\beta(\nu_5) = \beta(\nu_4)^\flat  = (p,\infty, p, \begin{smallpmatrix}
0 & 0 & 1 & \nopath\\
  \nopath  & \nopath & 1 & \nopath\\
  \nopath & \nopath & 0 & \nopath\\
  \nopath & \nopath & \nopath & 1
  \end{smallpmatrix})
\]
Observe that $\beta(\nu_5)$ is a witness of non-domination.
\end{runningexample}

\begin{figure}
         \centering
         \includegraphics[width=0.45\linewidth]{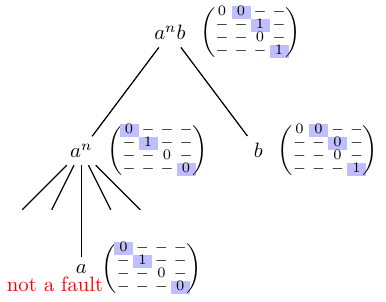}
         \caption{
         A factorisation tree for the word $a^nb$ with contributors highlighted.
         }
         \label{fig:treealt}
\end{figure}

\begin{runningexample}
Recall that contributors are defined top down, thus whether a node is a fault depends on the context in which it sits. Observe that $\nu_5$ of \cref{fig:tree}, which is a fault in that context, would not be fault if the tree were rooted at $\nu_3$.  This is because the contributors are different in this scenario, which is depicted in \cref{fig:treealt}. In this case the node corresponding to $a$ has a $1$ in an entry of the contributors, and is therefore not a fault, and does not induce a witness.
\end{runningexample}

Formally, every node $\nu$ in the tree is labelled by an element $\alpha(\nu)$ of $\overline{\ideal}_{\autA,\autB}$. We now associate nodes $\nu_2,\dots, \nu_m$ with elements of $\ideal_{\autA,\autB}$, which we denote by $\beta(\nu)$, defined by the following:
\begin{itemize}
\item If $h=2$, let $\beta(\nu_2) = \alpha(\nu_1)^\#$. This means we take stabilisation of the child's label.
\item If $\nu_h$ is a product of $\nu_{h-1}$ and $\nu'$, we let $\beta(\nu_h)= \beta(\nu_{h-1})\otimes\alpha(\nu')$.
\item If $\nu_h$ is a product of $\nu'$ and $\nu_{h-1}$, we let $\beta(\nu_h)= \alpha(\nu')\otimes\beta(\nu_{h-1})$.
\item  If $\nu_h$ is the idempotent product, with $\nu_{h-1}$ a middle child (that is, neither the left child nor the right) we let $\beta(\nu_h) = \beta(\nu_{h-1})^\flat$. This means we flatten the $\beta$-label of the child. 
\item If $\nu_h$ is the idempotent product, with $\nu_{h-1}$ as the left child (resp. right child), we let $\beta(\nu_{h}) = \beta(\nu_{h-1}) \otimes \alpha(\nu_{h-1})$ (resp. $\beta(\nu_{h}) = \alpha(\nu_{h-1}) \otimes \beta(\nu_{h-1})$).
\end{itemize}
Note we \emph{only} associate $\beta$-labels with nodes on the path $\nu_2,\dots, \nu_m$, and not other nodes in the tree.

We will observe that the $\beta$-labelling of the root, $\beta(\nu_m)$, is a tractable witness of non-domination. First, by construction, it has the correct shape: it is constructed with a single stabilisation at $\beta(\nu_2)$, and subsequently only products with elements of $\overline{\ideal}_{\autA,\autB}$ and nested flattening operations. Since the height of the tree is bounded by $3|\overline{\ideal}_{\autA,\autB}|$ then so to is the number of flattening operations defining $\beta(\nu_m)$. All is left is to show that it is a witness, which we do with the following property (in which we recall $C_{\nu}$ are the contributors of a node $\nu$).

\begin{prop}
\label{lemma:betanumiswitness}
For all $h\in\{2,\dots,m\}$, we have $\overline{\beta(\nu_h)} = \alpha(\nu_h)$ and if $\beta(\nu_h)= (p,x,q,M)$, we have $x=\infty$ and $M_{i,j} \le 1$ for all $(i,j) \in C_{\nu_h}$.
\end{prop}

Observe that this immediately implies that $\beta(\nu_m) \in \ideal_{\autA,\autB}$ is a witness.

\begin{proof}
We proceed by induction on $h$ starting from $h=2$.

\begin{case}[Base case, $h=2$]

Recall that $\beta(\nu_2) = \alpha(\nu)^\#$. Due to $\nu$ being an idempotent fault we can assume $\alpha(\nu)$ takes the form $(p,1,p,M)$, hence $\beta(\nu_2) = (p,\infty,p, N)$ where $N = M \otimes M' \otimes M$, and $M'$ replaces diagonal elements of $M$ with their stabilisation, i.e. $M'_{\ell,\ell}=(M_{\ell,\ell})^{\#}$. Since $\nu$ is a fault $M_{\ell,\ell} = 0$ for $(\ell,\ell) \in C_{\nu}$, hence $M'_{\ell,\ell} = 0$. Thus for $(i,j)\in C_{\nu_2}$ $M_{i,j} = \max_{\ell,k} M_{i,\ell}+ M'_{\ell,k} +  M_{k,j} \le 1$. Indeed for $(\ell,k)$ that could contribute to the maximum, $ M_{i,\ell}, M_{k,j} \le 1$ by definition, and either $\ell \neq k$ and $M'_{\ell,k} \le 1$, or $\ell = k$, $(\ell,k) \in C_{\nu}$ and hence, $M'_{\ell,k} = 0$. 
\end{case}

\begin{case}[$\nu_h$ is a product]
Let us assume $\nu_h$ is a product of $\nu_{h-1}$ and $\nu'$.
We then have $\overline{\beta(\nu_h)} = \overline{\beta(\nu_{h-1}) \otimes \alpha(\nu')} = \overline{\beta(\nu_{h-1})} \otimes \alpha(\nu') = \alpha(\nu_{h-1}) \otimes \alpha(\nu') = \alpha(\nu_{h})$.

By induction hypothesis,  $\beta(\nu_{h-1}) = (p,\infty,r,N')$ for some $p,r, N'$. Let us denote $\alpha(\nu') = (r,x,q,M)$ for some $q,M$ and some finite $x$. Then $\beta(\nu_h) = (p,\infty,q,N'\otimes M)$. Let $N = N'\otimes M$.

Consider $(i,j) \in C_{\nu_h}$. We have $N_{i,j} = \max_{\ell} N'_{i,\ell} + M_{\ell,j}$. Observe that $(i,\ell)\in C_{\nu_{h-1}}$ whenever $M_{\ell,j} \ne -\infty$ and $N'_{i,\ell} \ne -\infty$. Hence $N'_{i,\ell} \le 1$ by induction and $M_{\ell,j}\le 1$ by definition, since $\alpha(\nu) \in\overline{\ideal}_{\autA,\autB}$. Hence $N_{i,j} \le 1$.

The case $\nu_h$ is a product of some $\nu'$ and $\nu_{h-1}$ is similar.
\end{case}

\begin{case}[$\nu_h$ is idempotent such that $\nu_{h-1}$ is the left-most child]

First, note that we have $\alpha(\nu_h) = \alpha(\nu_{h-1})= \overline{\beta(\nu_{h-1})}$ by induction, and $\overline{\beta(\nu_{h})} = \overline{\beta(\nu_{h-1}) \otimes \alpha(\nu_{h})}$ by definition. Hence, $\overline{\beta(\nu_{h})} = \alpha(\nu_h)$ by idempotency. By induction hypothesis, $\beta(\nu_{h-1}) = (p,\infty,p,M)$, for some $p,M$, so $\beta(\nu_h)= (p,\infty,p,M\otimes \overline{M})$. 

Let $(i,j)$ in $C_{\nu_h}$. Then by definition of contributors, for all $\ell$ such that $\overline{M}_{i,\ell} \neq -\infty$ and $\overline{M}_{\ell,j} \neq -\infty$, we have $(i,\ell)$ in $C_{\nu_{h-1}}$.  Hence, by induction hypothesis, for such $\ell$, $M_{i,\ell} \leq 1$. We then have $(M\otimes \overline{M})_{i,j} = \max_{\ell} M_{i,\ell} + \overline{M}_{\ell,j} \leq 1$.
\end{case}

\begin{case}[$\nu_h$ is idempotent such that $\nu_{h-1}$ is the right-most child]
This case is symmetric to the previous one.
\end{case}

\begin{case}[$\nu_h$ is idempotent such that $\nu_{h-1}$ is a middle child]
By idempotence, note that $\alpha(\nu_h) = \alpha(\nu_{h-1})$. Also note that for idempotent $e\in \overline{\ideal}_{\autA,\autB}$ we have $e^\flat = e$, thus $\overline{e^\flat} = \overline{e} = e$. Therefore we have $\overline{\beta(\nu_{h})} = \overline{(\beta(\nu_{h-1}))^\flat} = \overline{\alpha(\nu_{h-1})^\flat} = \alpha(\nu_{h-1}) =  \alpha(\nu_h)$.

By induction hypothesis, $\beta(\nu_{h-1}) = (p,\infty,p,M)$,  for some $M$, and thus we have $\beta(\nu_{h})= (p,\infty,p,M)^\flat  = (p,\infty,p,M^\flat)$. Let $N = M^\flat$.

Consider $(i,j)\in C_{\nu_h}$.
We have $N_{i,j} = \max_{\ell,k} \overline{M}_{i,\ell} + \langle M^3 \rangle_{\ell,k} + \overline{M}_{k,j}$, where $\langle M^3 \rangle$ is the matrix $M^3$ where all the non diagonal elements are replaced by their barred version.
Observe that if $\ell \ne k$ then $\overline{M}_{i,\ell},\langle M^3 \rangle_{\ell,k}, \overline{M}_{k,j}$ cannot be $\infty$, as each entry has been replaced by its barred version. 

Thus it remains to verify $\overline{M}_{i,\ell} + \langle M^3 \rangle_{\ell,\ell} + \overline{M}_{\ell,j} \le 1$. We show that $\langle M^3 \rangle_{\ell,\ell}\le 1$ for $\ell$ such that $M_{i,\ell} \ne -\infty$ and $M_{\ell,j} \ne-\infty$. Note that $M^3_{\ell,\ell} \le 1$ if and only if $M_{\ell,s} + M_{s,t} + M_{t,\ell}\le 1$ for all $s,t$. For any such $\ell,s,t$ in which all three are not $-\infty$, the pairs $(\ell,s),(s,t)$ and $(t,\ell)$ are in the set of contributors $C_{\nu_{h-1}}$ of $\nu_{h-1}$. Thus by induction all three are less than or equal to $1$ and thus $M^3_{\ell,\ell} \le 1$.
\qedhere
\end{case}
\end{proof}

\section{Presence of witness implies non big-O}
\label{2implies1}
\setcounter{case}{0}

In this section, we assume that there is a witness of non-domination in  $\ideal_{\autA,\autB}$ and we construct a sequence of words $(w_i)_{i\in \mathbb{N}}$ such that for all positive integer $c$, there is $i$ such that: 
\[
\funcA(w_i) > c\cdot \funcB(w_i) + c.
\]

We will prove the following property:

\begin{prop}
\label{prop:words-construct}
For all $(p,x,q,M)$ in $\ideal_{\autA,\autB}$, for all $s\in \N$ there exists a pair $(w_s,x_s)$, with $w_s$ a word over $\Sigma^*$ and $x_s\in \N$ with the following properties: 
\begin{enumerate}[label={\normalfont(\arabic*)}]
\item $p \xrightarrow{w_s:x_s} q$ in $\autA$ with $\overline{x} = \overline {x_s}$,
\item $\overline{M_\autB(w_s)} = \overline{M}$,  
\item if $x = \infty$, for all $i,j$ such that $M_{i,j} \leq 1$, we have $x_s \geq s (M_\autB(w_s))_{i,j} + s$.
\end{enumerate}
\end{prop}

Note that applying this property to a witness of non-domination gives the expected result and concludes the proof.

\begin{proof}
We prove the proposition by structural induction on $\ideal_{\autA,\autB}$. Consider an element $(p,x,q,M)$ in $\ideal_{\autA,\autB}$. By definition of $\ideal_{\autA,\autB}$, $(p,x,q,M)$ is either a generator of $\ideal_{\autA,\autB}$ representing a letter, the product of two elements of $\ideal_{\autA,\autB}$, the stabilisation or the flattening of an element of $\ideal_{\autA,\autB}$.

\begin{case}[Base case: generator representing letters]
Consider an element $(p,\overline{y},q,\overline{M_\autB(a)})$ such that  $a\in \Sigma, p\xrightarrow{a:y}q \text{ in } \autA$. We associate with every $s$ the word $w_s = a$. Since $\overline{y} < \infty$, there is nothing to prove for (3).
\end{case}

\begin{case}[Product of two elements]

Suppose $(p,x,q,M) = (p,y,r,N)\otimes (r,z,q,P)$ with $(u_s,y_s)_{s\in\N}$ and $(v_s,z_s)_{s\in\N}$ given by induction. 
\begin{itemize}
\item If $y\le 1$ and $z\le 1$ (and so $x \le 1$), we define $w_s = u_sv_s$ and $x_s = y_s+z_s$. (1) and (2) are immediate by definition and there is nothing to prove for (3).

\item If $y=z=\infty$, we define $w_s = u_sv_s$ and $x_s = y_s+z_s$. (1) and (2) are immediate by definition. For (3), intuitively, since both $y=z=\infty$ we can straightforwardly bound both $y_s$ and $z_s$ through their respective words. Suppose $M_{i,j} \le 1$, we have $M_\autB(w_s)_{i,j} = M_\autB(u_s)_{i,\ell} + M_\autB(v_s)_{\ell,j}$ for some $\ell$.
Note that we have $N_{i,\ell} \le 1$ and $P_{\ell, j} \le 1$, otherwise $M_{i,j} = \infty$. Hence $s(M_\autB(u_s)_{i,\ell})+ s \le y_s$ and $s(M_\autB(v_s)_{\ell,j})+ s \le y_s$.
\begin{align*}
s(M_\autB(w_s)_{i,j}) + s & = s(M_\autB(u_s)_{i,\ell} + M_\autB(v_s)_{\ell,j} ) + s
\\ &\le  sM_\autB(u_s)_{i,\ell} + s +  sM_\autB(v_s)_{\ell,j} + s
\\&\le y_s + z_s  = x_s \quad \text{ as required.}
\end{align*}

\item If $y=\infty$ but not $z$, let $\Theta$ be the maximum value appearing in the matrix $M_\autB(v_0)$. We define $w_s = u_{s(\Theta+1)}v_0$ and $x_s = y_{s(\Theta+1)}+z_0$. (1) and (2) are immediate by definition. For (3), we will only be able to use property (3) inductively from $y=\infty$ but not $z$, thus we only use the short word $v_0$ (to ensure path compatibility) with the sufficiently larger word $u_{s(\Theta+1)}$. Suppose $M_{i,j} \le 1$, we have $M_\autB(w_s)_{i,j} = M_\autB(u_{s(\Theta+1)})_{i,\ell} + M_\autB(v_0)_{\ell,j}$ for some $\ell$.
Note that we have $N_{i,\ell} \le 1$, otherwise $M_{i,j} = \infty$. Hence, 
$$s(\Theta+1)(M_\autB(u_{s(\Theta+1)})_{i,\ell})+ {s(\Theta+1)} \le y_{s(\Theta+1)}.$$ 
\begin{align*}
s(M_\autB(w_s)_{i,j}) + s & = s(M_\autB(u_{s(\Theta+1)})_{i,\ell} + M_\autB(v_0)_{\ell,j} ) + s
\\ &\le sM_\autB(u_{s(\Theta+1)})_{i,\ell} + s\Theta +s  \tag{since $M_\autB(v_0)_{\ell,j} \le \Theta$}
\\ &\le s(\Theta+1)M_\autB(u_{s(\Theta+1)})_{i,\ell} +s(\Theta+1)
\\ &\le y_{s(\Theta+1)} \le y_{s(\Theta+1)} + z_0 = x_s.
\end{align*}

\item The case of $z=\infty$ but not $y$ is symmetric.
\end{itemize}

\end{case}
\begin{case}[Stabilisation of an element]

Suppose $(p,x,p,M) = (p,y,p,P)^\sharp$ and $(u_s,y_s)_{s\in\N}$ given by induction. 
\begin{itemize}
\item If $y = 0$ (and hence $x=0$), let $w_s = u_s$ and $x_s=y_s$. (1) and (2) are immediate by definition - since $\overline{P^\sharp} = \overline{M}$ as $P$ is path-idempotent - and there is nothing to prove for (3). 
\item If $y=\infty$ (and hence $x=\infty$), let $w_s = u_s$ and $x_s=y_s$. (1) and (2) are immediate by definition. Observe that if $M_{i,j} \le 1$ then $P_{i,j} \le 1$, thus 
\[
s(M_\autB(w_s)_{i,j}) + s = s(M_\autB(u_s)_{i,j}) + s \le y_s = x_s.
\]
\item If $y=1$ (and hence $x=\infty$), let $\Theta$ be the maximum value appearing in the matrix $M_\autB(u_0)$ and recall $|Q_\autB|$ is the number of states of $\autB$.
We define $w_s = u_0^{s(\Theta|Q_\autB| + 1)}$ and $x_s = s_0s(\Theta|Q_\autB| + 1)$. (1) and (2) are immediate by definition. For (3), intuitively if $M_{i,j} \leq 1$, then the repetition of $u_0$ cannot access a positive cycle between $i$ and $j$, hence bounding the weight of $M_\autB(w_s)_{i,j}$, while iterating $u_0$ sufficiently many times will make $x_s$ as large as needed. 
Formally, if $M_{i,j} \leq 1$, then for all $\ell$ such that $P_{i,\ell}$ and $P_{\ell,j}$ are both different from $-\infty$, we have $P_{\ell,\ell} = 0$. Hence, since $P$ is path-idempotent and $\overline{M_\autB(u_0)} = \overline{P}$ by induction, $M_\autB(w_s)_{i,j}$ has value at most $\Theta|Q_\autB|$. On the other hand, the weight of $w_s$ in $\autA$ from $p$ to $p$ is at least $s(\Theta|Q_\autB| + 1)$, since $y=1$. Hence,
\[
s(M_\autB(w_s)_{i,j}) + s \le s\Theta|Q_\autB| + s \le x_s.
\]

\end{itemize}
\end{case}
\begin{case}[Flattening of an element]

Suppose $(p,x,p,M) = (p,y,p,P)^\flat$ and $(u_s,y_s)_{s\in\N}$ given by induction. 
\begin{itemize}
\item If $x=y\le 1$ let $w_s = u_s$ and $x_s=y_s$. (1) and (2) are immediate by definition -- since $\overline{P^\flat} = \overline{M}$ as $P$ is path-idempotent -- and there is nothing to prove for (3). 

\item Otherwise, we have $x= y=\infty$. 
Let $\Theta_s$ be the maximum value appearing in the matrix $M_\autB(u_s)$, $|Q_\autB|$ the number of states of $\autB$ and $K_s = |Q_\autB|\Theta_s + 1$. We define $w_s = (u_s)^{K_s}$ and $x_s = K_s y_s$.
(1) and (2) are immediate by definition. We prove (3).

By induction, if $P_{i,j} \le 1$ then $s( M_\autB(u_s)_{i,j}) + s \le y_s$. Let $R_s = \displaystyle\max_{i,j : P_{i,j}\le 1} M_\autB(u_s)_{i,j}$. In particular, 
\begin{equation}\label{eq:sr}sR_s + s \le y_s.\end{equation}

Consider $i,j$ such that $M_{i,j}\leq 1$. Since $P$ is path-idempotent and by definition of flattening, necessarily for all $\ell$ such that both $P_{i,\ell}$ and $P_{\ell,j}$ are different from $-\infty$, we have $P_{\ell,\ell} \le 1$ ($\star$).

We have:
\begin{align}M_\autB(w_s)_{i,j} &= M_\autB(u_s^{K_s})_{i,j}  \notag
\\ &= \max_{\substack{i_0,i_{1},i_2,\dots,i_{K_s}\\i_{0}=i,\ i_{K_s}=j }}  M_\autB(u_s)_{i_0,i_1} + M_\autB(u_s)_{i_1,i_2} + \dots + M_\autB(u_s)_{i_{K_s-1},i_{K_s}} \label{eq:boundflatteningijpath}
\end{align}

Consider the path $i_0,i_1,\dots,i_{K_s}$ that achieves the maximum in \cref{eq:boundflatteningijpath}. Since we have $K_s\ge|Q_\autB|$, there exists $n<m$ such that $i_n= i_m$ and all elements $i_0,\dots,i_{n},i_{m+1}\dots,i_{K_s}$ are distinct. By ($\star$), we have $P_{i_n,i_n} \le 1$ and furthermore, $P_{i_\ell,i_{\ell+1}} \le 1$ for $n\le \ell \le m-1$ since $P_{i_n,i_m}\le 1$. By definition of $R_s$, we then have $M_\autB(u_s)_{i_{\ell},i_{\ell+1}} \le R_s$. Hence all components of \cref{eq:boundflatteningijpath}, \emph{except} those between $i_0,\dots,i_n$ and $i_m,\dots,i_{K_s}$, are bounded above by $R_s$, and the remaining, of which there are at most $|Q_\autB|$, are bounded above by $\Theta_s$.
We have: 
\begin{equation}\label{eq:boundflatteningMwsij}M_\autB(w_s)_{i,j} \le |Q_\autB|\Theta_s + K_s R_s.\end{equation}
So, we have:
\begin{align*}
s(M_\autB(w_s)_{i,j}) + s &\le s( |Q_\autB|\Theta_s + K_s R_s) + s \tag{by \cref{eq:boundflatteningMwsij}}
\\&= s(K_s R_s +  |Q_\autB|\Theta_s   +1)
\\&= s(K_s R_s +  K_s)
\tag{by choice of $K_s$}
\\&= K_s (sR_s +  s)
\\&\le K_s y_s  \tag{by \cref{eq:sr}}
\\& = x_s. \tag*{\qed}
\end{align*}
\end{itemize}
\end{case}
\renewcommand{\qed}{}
\end{proof}

\begin{runningexample}
We compute the sequence $w_s$ for the nodes inducing the tractable witness in our example:
\begin{itemize}
\item The leaves, labelled by $a$ and $b$, induce the sequences $w_s = a$ for all $s$, and $w_s = b$ for all $s$ respectively.
\item $\beta(\nu_2)$ is generated as the stabilisation of an element with word $u_s = a$ in which $y = 1$ (Case 3.3), hence $w_s = a^{s(\Theta |Q_\autB| + 1)} = a^{5s}$, where $\Theta =\max_{i,j} M_\autB(a)_{i,j} = 1$ and $|Q_\autB| = 4$.
\item $\beta(\nu_3)$ is the product of elements with $u_s = a^{5s}$ and $v_s = b$, where $y= \infty$, but not $z$, and so we have $w_s = u_{s(\Theta+1)}v_0 = a^{10s}b $, where $\Theta  =\max_{i,j} M_\autB(b)_{i,j} = 1$.
\item $\beta(\nu_4)$ is the product of elements with $u_s = ab$ and $v_s = a^{10s}b$, where $z= \infty$ but not $y$, and so we have $w_s = ab v_{s(\Theta+1)}= ab a^{20s}b$, as $\Theta =\max_{i,j} M_\autB(ab)_{i,j} = 1$.
\item The tractable witness $\beta(\nu_5)$ is the flattening of an element with $u_s = aba^{20s}b$, where $y= \infty$, thus (by Case 4.2) $w_s = u_s^{|Q_\autB|\Theta + 1} = (aba^{20s}b)^{4\cdot 20s + 1} $, where $\Theta  =\max_{i,j} M_\autB(u_s)_{i,j} =  20s$.
\end{itemize}
Hence, our witness shows that for every $s$, $w_s=(aba^{20s}b)^{80s + 1}$ is a contradiction to $\funcA \le s \funcB + s$. Apart from the additional complexity introduced by the constants, the sequence matches our expectations from \cref{re:notbig-O,re:actualtractablewitness}.

Indeed, $\funcA(w_s) = (20s+3)(80s+1) = 800s^2 +260s+3$ increases quadratically in $s$, while $\funcB(w_s) = 160s + 2$, maximised by counting $b$'s, only increases linearly in $s$, and in particular:
\[
800s^2 +260s+3 > s(160s+2) + s \text{ for every }s \in\mathbb{N}.\qedhere
\]
\end{runningexample}

\section{Max-Plus automata with increasingly complex witnesses}
\label{sec:example}

We have defined tractable witnesses for two max-plus automata $\autA$ and $\autB$ as a means to decide whether $\autA$ is big-O of $\autB$. One of the specific characteristics of tractable witnesses is the number of nested $\flat$ operations appearing in the expression representing the witness. In this section, we construct a sequence of pairs of max-plus automata $(\autA_n, \autB_n)_{n\geq 1}$ with $\autA_n$ not big-O of $\autB_n$ such that (1) there is a tractable witness between $\autA_n$ and $\autB_n$ that has $n-1$ nested $\flat$ operations, but (2) there is no tractable witness with strictly less than $n-1$ nested $\flat$ operations. We go even further by defining the $\sharp\flat$-height of an element of $\ideal_{\autA,\autB}$ (essentially the minimal number of nested $\sharp$ and $\flat$ operations required to produce this element from the basic elements corresponding to the letters and product) and we prove that for $\autA_n$ and $\autB_n$ there is no witness in $\ideal_{\autA_n,\autB_n}$ that are of $\sharp\flat$-height less than $n$. 

Recall that, for $\autA$ deterministic, we have defined $\ideal_{\autA,\autB}$ as generated by the set 
$$X_{\autA, \autB} = \{(p,\overline{x},q,\overline{M_\autB(a)}) \mid a\in \Sigma, p\xrightarrow{a:x}q \text{ in } \autA\}$$ 
and closed under product, stabilisation and flattening operations. We now define $\sharp\flat$-expressions, these describe how an element of $\ideal_{\autA,\autB}$ is generated from $X_{\autA, \autB}$.

\begin{defi}
The set of $\sharp\flat$-expressions over $X_{\autA, \autB}$ is defined by induction as the minimal set such that:
\begin{itemize}
\item for all $x \in X_{\autA, \autB}$, $x$ is a $\sharp\flat$-expression,
\item if $s$ and $t$ are $\sharp\flat$-expressions, then $st$ is a $\sharp\flat$-expression,
\item if $s$ is a $\sharp\flat$-expression, then $s^\sharp$ is a $\sharp\flat$-expression,
\item if $s$ is a $\sharp\flat$-expression, then $s^\flat$ is a $\sharp\flat$-expression.
\end{itemize} 
\end{defi}

We map every $\sharp\flat$-expressions over $X_{\autA, \autB}$ to the element in $\ideal_{\autA,\autB}\cup \{\bot\}$ corresponding to doing the product, $\sharp$ and $\flat$-operations as in the expression. Additionally, the expression will be mapped to $\bot$ if a $\sharp$ and $\flat$ operations are applied to elements which are not path-idempotent. Formally, we define $\pi$ the following projection from the set of $\sharp\flat$-expressions over $X_{\autA,\autB}$ to $\ideal_{\autA,\autB}\cup \{\bot\}$ by induction:
\begin{itemize}
\item if $x \in X_{\autA,\autB}$, then $\pi(x) = x$,
\item if $s$ and $t$ are $\sharp\flat$-expressions over $X$, then $\pi(st) = \pi(s)\otimes \pi(t)$,
\item if $s$ is a $\sharp\flat$-expression over $X$, then $\pi(s^\sharp) = \pi(s)^\sharp$ if $\pi(s)$ is path-idempotent, and $\bot$ otherwise,
\item if $s$ is a $\sharp\flat$-expression over $X$, then $\pi(s^\flat) = \pi(s)^\flat$ if $\pi(s)$ is path-idempotent, and $\bot$ otherwise.
\end{itemize}

\newcommand{\sfh}[1]{\sharp\flat\text{-height}(#1)}

\begin{defi}
The $\sharp\flat$-height of a $\sharp\flat$-expression, denoted $\sfh{\cdot}$ is defined by induction as follows:
\begin{itemize}
\item for all $x \in X_{\autA,\autB}$, $\sfh{x} = 0$,
\item if $s$ and $t$ are $\sharp\flat$-expressions, then $\sfh{st} = \max(\sfh{s},\sfh{t})$,
\item if $s$ is a $\sharp\flat$-expression, then $\sfh{s^\sharp} = 1+ \sfh{s}$,
\item if $s$ is a $\sharp\flat$-expression, then $\sfh{s^\flat} = 1+ \sfh{s}$,
\end{itemize} 
\end{defi}

\begin{defi}
\label{defi:sfheightelement}
The $\sharp\flat$-height of an element $z$ of $\ideal_{\autA,\autB}$ is defined as 
$$\sharp\flat\text{-height}(z) = \min\{\sfh{s}\mid\pi(s)=z\}.$$
\end{defi}

Consider a tractable witness of non-domination $g\in \ideal_{\autA,\autB}$; which can be described by a $\sharp\flat$-expressions of the form
\begin{equation*}\label{eq:sftractable}g_0 (g_1 (\dots (g_{k-2} ( g_{k-1} (g_{k})^\#g'_{k-1})^\flat g'_{k-2})^\flat \dots)^\flat g'_{1})^\flat g'_{0}.\end{equation*} 
The $\sharp\flat$-height of this expression is $k$, which shows that the $\sharp\flat$-height of $g$ is \textbf{at most} $k$; but there may be another $\sharp\flat$-expression for $g$ with lower height. The following proposition shows that allowing general witnesses does not help in terms of $\sharp\flat$-height and tractable witnesses have $\sharp\flat$-height as low as can be.

\begin{prop}
\label{prop:witnesssfh}
For all max-plus automata $\autA$ and $\autB$ (with $\autA$ deterministic), if there exist a witness of $\sharp\flat$-height $n$ for some $n \in \N$ then there exist a tractable witness of $\sharp\flat$-height at most $n$.
\end{prop}

\begin{proof}
We will say that an element of $\ideal_{\autA,\autB}$ has a tractable witness shape if it can be written as:
\[g_0 (g_1 (\dots (g_{k-2} ( g_{k-1} (g_{k})^\#g'_{k-1})^\flat g'_{k-2})^\flat \dots)^\flat g'_{1})^\flat g'_{0},\] 
for some $k \geq 0$, where all the $g_i$ and $g'_i$ are products of elements in $X_{\autA,\autB}$.

We prove by induction that for all elements $(p,\infty,q,M)$ of $\ideal_{\autA,\autB}$ of $\sharp\flat$-height $n$, there is an element $(p,\infty,q,M')$ with a tractable witness shape and $\sharp\flat$-height at most $n$ such that $M' \leq M$ (in a component-wise meaning, where $-\infty<0<1<\infty$). Applying this to a witness leads to the expected result.

Let $z$ be an element $(p,\infty,q,M)$ of $\ideal_{\autA,\autB}$ of $\sharp\flat$-height $n$.
\begin{itemize}
\item If $z$ is in $X_{\autA,\autB}$, then $z$ has a tractable witness shape and is of $\sharp\flat$-height $n=0$.
\item If $z$ is equal to $z_1z_2$, then $z_1= (p,x_1,r,M_1)$, $z_2 = (r,x_2,q,M_2)$, $z_1$ and $z_2$ have $\sharp\flat$-height at most $n$, $M=M_1\otimes M_2$, and one of $x_1$ or $x_2$ (or both) is equal to $\infty$. Let's assume $x_1 = \infty$, the other case is similar. By induction hypothesis on $z_1$, there is an element $z'_1 = (p,\infty,r,M'_1)$ with a tractable witness shape and $\sharp\flat$-height at most $n$ such that $M'_1 \leq M_1$. We deduce that the element $z'_1\overline{z_2} = (p,\infty,q,M'_1\otimes \overline{M_2})$ has a tractable witness shape of $\sharp\flat$-height at most $n$ and $M'_1\otimes \overline{M_2} \leq M$. In this context, by $\overline{z_2}$ we mean the element of $\ideal_{\autA,\autB}$ generated by $\sharp\flat$-expression for $z_2$ in which the stabilisation and flattening operations are omitted. 
\item  If $z$ is equal to $z_1^\sharp$, then if $z_1= (p,\infty,q,M_1)$, we can directly apply the induction hypothesis to $z_1$ and immediately get an element satisfying the conditions (without needing to $\sharp$ it). If $z_1= (p,1,q,M_1)$, then $\overline{z_1}^\sharp = (p,\infty,q,\overline{M_1}^\sharp)$ satisfies the condition: it has a tractable witness shape of $\sharp\flat$-height $1$, so at most $n$ and $\overline{M_1}^\sharp \leq M_1^\sharp \leq M$.
\item If $z$ is equal to $z_1^\flat$, then $z_1= (p,\infty,q,M_1)$ and has $\sharp\flat$-height at most $n-1$. We can apply the induction hypothesis to $z_1$. There is an element $z'_1 = (p,\infty,q,M'_1)$ with a tractable witness shape of $\sharp\flat$-height at most $n-1$ such that $M'_1 \leq M_1$. Hence, ${z'_1}^\flat = (p,\infty,q,{M'_1}^\flat)$ has a tractable witness shape of $\sharp\flat$-height at most $n$ with ${M'_1}^\flat \leq {M_1}^\flat \leq M$.  \qedhere
\end{itemize}
\end{proof}

\begin{rem}
Our main result, \cref{thm:4assertion}, asserts four equivalencies which are proven in a cycle,
    \cref{prop:witnesssfh} directly proves that (2) implies (3) in \cref{thm:4assertion}.
\end{rem}

\begin{thm}
There exists a sequence of pairs of max-plus automata $(\autA_n, \autB_n)_{n \geq 1}$ such that for all $n \geq 1$,
\begin{enumerate}
\item there exists a tractable witness in $\ideal_{\autA_n,\autB_n}$ of $\sharp\flat$-height $n$,
\item there is no witness in $\ideal_{\autA_n,\autB_n}$ of $\sharp\flat$-height strictly less than $n$.
\end{enumerate}
\end{thm}

In the rest of the paper, we fix a positive integer $n$. We give the construction of the max-plus automata $\autA_n$ and $\autB_n$ in Section~\ref{section:constructionAB}, and prove items 1. in Section~\ref{section:existence} and 2. in Section~\ref{section:nonexistence}.

For the rest of the paper, $\Sigma_n = \{a_1, a_2, \ldots, a_n\}$ denotes an alphabet with $n$ letters\footnote{The following construction would also work with a two letter alphabet, by decomposing each transition into $O(log(n))$ successive transitions in a standard way.}. 

\subsection{Construction of \texorpdfstring{$\autA_n$}{A\_n} and \texorpdfstring{$\autB_n$}{B\_n}}
\label{section:constructionAB}

The max-plus automata $\autA_n$ and $\autB_n$ are both defined on alphabet $\Sigma_n$.
The automaton $\autA_n$ is defined as the deterministic max-plus automata which computes the length of the input word, and we denote by $q$ its only state. 

\begin{center}
   \begin{tikzpicture} [node distance = 1.5cm, auto, scale = 1.7] 
\tikzstyle{vertex}=[circle, fill=black!25,minimum size=20pt,inner sep=1pt ] 
    \node (1) [vertex] {$q$};
    \node (1f) [right of =1] {};
    \node (nom) [] at (-2,0.25) {$\mathcal{A}_{n} : $};
    \node (balanceL) [] at (-2.5,0) {};
    \node (balanceR) [] at (2.5,0) {};
    \path [->,>=stealth', semithick] 
        node [left of =1]{} edge node {\footnotesize$0$} (1) 
        (1) edge node {\footnotesize$0$} (1f)
        (1) edge [loop above] node {$a_1,...,a_n : 1$}();
\end{tikzpicture} 
\end{center}

The automaton $\autB_n$ is defined as the maximum (disjoint union) of $n$ max-plus automata $\autB_{n,1}, \autB_{n,2}, \ldots \autB_{n,n}$. Automaton $\autB_{n,1}$ has one state, and all other $\autB_{n,m}$ have $3$ states each (hence $\autB_n$ has $3n -2$ states).

\begin{center}
   \begin{tikzpicture} [node distance = 1.5cm, auto, scale = 1.7] 
\tikzstyle{vertex}=[circle, fill=black!25,minimum size=20pt,inner sep=1pt ] 
    \node (1) [vertex] {$q_1$};
    \node (1f) [right of =1] {};
    \node (jsp) [] at (0,1) {$a_1:0$};
    \node (nom) [] at (-2,0.25) {$\mathcal{B}_{n,1} : $};
    \node (balanceL) [] at (-2.5,0) {};
    \node (balanceR) [] at (2.5,0) {};

    \path [->,>=stealth', semithick] 
        node [left of =1]{} edge node {\footnotesize$0$} (1) 
        (1) edge node {\footnotesize$0$} (1f)
        (1) edge [loop above] node {$a_2,...,a_n : 1$}();
\end{tikzpicture} 
\end{center}

For all $1< m \leq n$, $\mathcal{B}_{n,m}$:

\begin{center}
    \begin{tikzpicture} [node distance = 1.5cm, auto, scale = 2.5] 
\tikzstyle{vertex}=[circle, fill=black!25,minimum size=25pt ,inner sep=0pt ] 
    \node (1) [vertex] {$q_{m,1}$};
    \node (2) [vertex] at (2,0) {$q_{m,2}$};
    \node (3) [vertex] at (4,0) {$q_{m,3}$};
    \node (br2) [below right of=2] {};
    \node (3f) [right of =3] {};
    \node (2a) [] at (2,0.9) {$\Sigma_n \setminus \{ a_1, a_m \}:0$};

    \path [->,>=stealth', semithick] 
        node [left of =1]{} edge node {\footnotesize$0$} (1)
        node[below left of=2] {} edge node[below,yshift=0.1cm,xshift=0.1cm] {\footnotesize$0$} (2)
        (2) edge node[below,xshift=-0.1cm,yshift=0.1cm] {\footnotesize$0$} (br2)
        (3) edge node {\footnotesize$0$} (3f)
        (1) edge node {$a_m : 0$} (2)
        (2) edge node {$a_m : 0$} (3)
        (3) edge [bend left ]node {$a_{m+1},...,a_n : 0$} (2)
        (3) edge [bend left = 50]node {$a_{m+1},...,a_n : 0$} (1)
        (2) edge [bend left ]node {$a_{m+1},...,a_n : 0$} (1)
        (2) edge [loop above] node {$a_1 : 1$}()
        (3) edge [loop above] node {$a_1,...,a_m : 0$}()
        (1) edge [loop above] node {$\Sigma_n : 0$}();
        
\end{tikzpicture}
\end{center}

We define $\mathcal{B}_n$ as $\max(\mathcal{B}_{n,1}, \mathcal{B}_{n,2}, ..., \mathcal{B}_{n,n})$. By definition, the matrix associated to $\mathcal{B}_n$ for each word is a block diagonal matrix: the first block is of dimension $1 \times 1$ (for $\autB_{n,1}$), and all other blocks are of dimension $3\times 3$ (for $\autB_{n,m}$, $2 \leq m \leq n$). Given a word $w$, we will denote by $M_1(w)$, $M_2(w)$,... $M_n(w)$ these blocks (matrices), omitting $n$ which is now fixed.  

\begin{runningexample}
Note that $\mathcal{A}_n,\mathcal{B}_n$ generalises the running example, that is, when $n = 2$, $\mathcal{A}_n,\mathcal{B}_n$ are exactly the $\mathcal{A},\mathcal{B}$ from \cref{eg:running}. 
\end{runningexample}

Let us give a rough intuition of what automaton $\mathcal{B}_n$ is computing on words of the shape: 
$$(a_n(a_{n-1}(\ldots (a_2(a_1^{\ell_1})a_2)^{\ell_2} \ldots a_{n-1})^{\ell_{n-1}}a_n)^{\ell_n}$$
\begin{itemize}
\item $\mathcal{B}_1 = \mathcal{B}_{1,1}$ computes the function that maps each word to $0$.
\item $\mathcal{B}_2 = \max(\mathcal{B}_{2,1}, \mathcal{B}_{2,2}) = \mathcal{B}$ from the running example (up to renaming of letters) and computes the function that maps each word to the maximum between the longest block of $a_1$'s and the number of $a_2$'s.
\item $\mathcal{B}_3 = \max(\mathcal{B}_{3,1}, \mathcal{B}_{3,2}, \mathcal{B}_{3,3})$ computes a function that maps each word of the shape $(a_3(a_2(a_1^{\ell_1})a_2)^{\ell_2}a_3)^{\ell_3}$ 
to the maximum of the three following components:
\begin{itemize}
    \item (for $\mathcal{B}_{3,1}$) the number of $a_2$'s and $a_3$'s, so big-O of $\ell_2\ell_3$,
    \item (for $\mathcal{B}_{3,2}$) between all pairs of consecutive $a_3$'s, take the largest number of $a_1$'s appearing between two $a_2$'s, and then take the sum, so it would give big-O of $\ell_1\ell_3$,
    \item (for $\mathcal{B}_{3,3}$) the largest number of $a_1$'s appearing between two $a_3$'s (ignoring $a_2$'s), so big-O of $\ell_1\ell_2$.
\end{itemize}
\item for the general case, $\mathcal{B}_n$ computes a function that maps each word of the shape 
$$(a_n(a_{n-1}(\ldots (a_2(a_1^{\ell_1})a_2)^{\ell_2} \ldots a_{n-1})^{\ell_{n-1}}a_n)^{\ell_n}$$ 
to the maximum of the $n$ following components:
\begin{itemize}
    \item (for $\mathcal{B}_{n,1}$) the number of $a_2$'s, $a_3$'s,... and $a_n$'s, so big-O of $\ell_2\ell_3 \dotsm \ell_n$,
    \item (for $\mathcal{B}_{n,m}$ with $1<m\leq n$) between all pairs of consecutive letters $a_i$'s and $a_j$'s with $i,j > m$, take the largest number of $a_1$'s appearing between two $a_m$'s (ignoring the other letters), and then take the sum, so it would give big-O of $\ell_1\ell_2 \dotsm \ell_{m-1}\ell_{m+1}\dotsm\ell_{n}$.
\end{itemize}
\end{itemize}

\subsection{Existence of tractable witnesses of \texorpdfstring{$\sharp\flat$}{\#flat}-height \texorpdfstring{$n$}{n}}
\label{section:existence}

For $i=1, \ldots, n$, let $\boldsymbol{a_i}$ denote the element $(q,1,q,\overline{M_{\autB_n}(a_i)})$ corresponding to the letter $a_i$ in $\ideal_{\autA_n,\autB_n}$.

\begin{prop}
\label{prop:sfheightwitnessexists}
The element of $\ideal_{\autA_n,\autB_n}$ \[ \boldsymbol{w} = (\boldsymbol{a_n} (\boldsymbol{a_{n-1}} (\dots (\boldsymbol{a_{3}} ( \boldsymbol{a_{2}} (\boldsymbol{a_{1}})^\#\boldsymbol{a_{2}})^\flat \boldsymbol{a_{3}})^\flat \dots)^\flat \boldsymbol{a_{n-1}})^\flat \boldsymbol{a_{n}})^\flat\]  is a tractable witness of non-domination between $\autA_n$ and $\autB_n$ of $\sharp\flat$-height $n$.
\end{prop}

\begin{proof}
First of all, it is immediate by definition that $\boldsymbol{w} = (q,\infty,q, M)$ for some $M$. We are left to prove that all the coefficients in $M$ corresponding to a path from an initial state to a final state in $\autB_n$ are not $\infty$. These coefficients are $(M_1)_{1,1}$, $(M_m)_{1,2}$, $(M_m)_{1,3}$, $(M_m)_{2,2}$  and $(M_m)_{2,3}$ for $2\leq m \leq n$.

We have $M_{1}(a_1)=0$, and hence $(M_{1})_{1,1} \neq \infty$.

For $2 \leq m \leq n$ and $2 \leq i < m$, we have : 
  \begin{center}
      $M_{m}(a_1) =  
      \begin{pmatrix}
          0 & - & - \\
          - & 1 & -\\
          - & - & 0
      \end{pmatrix}$, 
      $M_{m}(a_i) = 
      \begin{pmatrix}
          0 & - & - \\
          - & 0 & - \\
          - & - & 0
      \end{pmatrix}$, and
      $M_{m}(a_m) = 
      \begin{pmatrix}
          0 & 0 & - \\
          - & - & 0 \\
          - & - & 0
      \end{pmatrix}$, 
\end{center}

which gives (path-idempotency is satisfied at every step):

\begin{center}
      $M_{m}(a_1)^\sharp = 
      \begin{pmatrix}
          0 & - & - \\
          - & \infty & -\\
          - & - & 0
      \end{pmatrix}$ and
      $(M_{m}(a_{m-1})(...)^\flat M_{m}(a_{m-1}))^\flat= 
      \begin{pmatrix}
          0 & - & - \\
          - & \infty & -\\
          - & - & 0
      \end{pmatrix}$.
  \end{center}

Finally,
    \begin{center}
      $M_{m}(a_m)(...)^\flat M_{m}(a_m) = 
      \begin{pmatrix}
          0 & 0 & \infty \\
          - &- & 0\\
          - & - & 0
      \end{pmatrix}$ and
      $(M(a_{m})_{m}(...)^\flat M(a_{m})_{m})^\flat= 
      \begin{pmatrix}
          0 & 0 & 1 \\
          - & - & 0\\
          - & - & 0
      \end{pmatrix}$.
  \end{center} 

Finally, using only products and $\flat$ operations (but no $\sharp$ operation) on matrices that do not contain any $\infty$ cannot create $\infty$. Hence, 
$$(M(a_n)_{m} (\ldots (M(a_{m})_{m}(...)^\flat M(a_{m})_{m})^\flat \ldots)^\flat M(a_n)_{m})^{\flat}$$
does not contain any $\infty$. This is true for all $2 \leq m \leq n$, and thus all
$(M_m)_{1,2}$, $(M_m)_{1,3}$, $(M_m)_{2,2}$  and $(M_m)_{2,3}$ are different from $\infty$, giving the expected result.
\end{proof}

\begin{runningexample}
\cref{prop:sfheightwitnessexists} entails that $(e_b(e_a)^\sharp e_b)^\flat$ is also a witness of non-domination for $\autA,\autB$.
\end{runningexample}

\subsection{Non-existence of witnesses of \texorpdfstring{$\sharp\flat$}{\#♭}-height strictly less than \texorpdfstring{$n$}{n}}
\label{section:nonexistence}

Using Propositions~\ref{prop:witnesssfh}, it is enough to prove that there is no tractable witness of $\sharp\flat$-height $n-1$ in $\ideal_{\autA_n,\autB_n}$ to obtain the expected result. Note that, if there were a tractable witness of  $\sharp\flat$-height $m < n-1$, then there would also exist a tractable witness of  $\sharp\flat$-height $n-1$: for any matrix $M$, $(M^\sharp)^\flat \leq M^\sharp$ then by repeatedly using the $\flat$ operation after the (only) inner-most $\sharp$ operation we could get a tractable witness of any height higher than $m$.

\begin{prop}
\label{prop:last}
There is no tractable witness of $\sharp\flat$-height $n-1$ in $\ideal_{\autA_n,\autB_n}$.
\end{prop}

Let $g$ be an element of $\ideal_{\autA_n,\autB_n}$ with the shape of a tractable witness, \[g = g_n (g_{n-1} (\dots (g_{2} (g_{1})^\#g'_{2})^\flat \dots)^\flat g'_{n-1})^\flat g'_{n},\] 
of $\sharp\flat$-height $n-1$, where each $g_i$ and $g'_i$ is a product (possibly empty, except for $g_1$) of some elements $\boldsymbol{a_j}$ for $i$ ranging between $1$ and $n$. We say that $g_i$ has $\boldsymbol{a_j}$ as factor if it appears in it.

Let us first start with the intuition. In any tractable witness,  $g_1$ must have $\boldsymbol{a_1}$ as its only factor, otherwise, $\autB_{n,1}$ will keep up with $\autA_n$ and it cannot be a witness. By using $\boldsymbol{a_1}$ in $g_1$, we have introduced $\infty$ into the matrix for every $\autB_{n,m}$ ($m\ne 1$) and must remove this $\infty$ by a sequence of flattening operations. Essentially (although not exactly), the flattening operation removes $\infty$'s that are not on the diagonal. Therefore the $g_i,g_i'$ ($i\ge 2$) in the witness  must be used to take $\infty$'s off of the diagonal, so that flattening removes the $\infty$'s. However, the $\autB_{n,m}$ are constructed in such a way that any $g_i, g_i'$ that removes $\infty$'s in $\autB_{n,m}$ cannot also remove $\infty$'s in any other $\autB_{n,m'}$ ($m'\ne m$), resulting in the need for at least $n-1$ flattening operations; one for each $B_{n,m}$. We now prove the result formally, in which we must carefully keep track of the locations of each $\infty$.

Let $g = (q,x,q,M)$ for some $x$ and $M$. We are going to prove that $g$ cannot be a (tractable) witness. If $x \neq \infty$, the result is immediate. Otherwise, we denote by $M^{(1)}$, $M^{(2)}$,... $M^{(n)}$ the diagonal blocks (matrices) of $M$ of size $1\times 1$ for $M^{(1)}$ and $3\times 3$ for $M^{(m)}$, $m\neq 1$ corresponding to the behaviours in the automata $\autB_{n,1}, \autB_{n,2}, \ldots \autB_{n,n}$, and by $P^{(m)}_i,P'^{(m)}_i$ the matrices corresponding to the $g_i$, $g'_i$ such that:

$$M^{(m)} = P^{(m)}_n (P^{(m)}_{n-1} (\ldots (P^{(m)}_2 (P^{(m)}_1)^\sharp P'^{(m)}_2)^\flat \ldots)^\flat P'^{(m)}_{n-1})^\flat P'^{(m)}_n$$

Note that it is enough to prove that, either $M^{(1)} = (\infty)$ or there exist $i$ between $2$ and $n$ such that one of $M^{(m)}_{1,2}$, $M^{(m)}_{2,2}$, $M^{(m)}_{1,3}$ or $M^{(m)}_{2,3}$ is $\infty$.

To prove this, we use the following three lemmas, their proofs follow in the following sections.

\begin{lem}
\label{lemma:one}
Let $1< m\leq n$. If for all $1\leq i < m$, $g_ig'_i$ has only factors $\boldsymbol{a_j}$ for $j\leq i$ then:
$$(P^{(m)}_{m-1} (\ldots (P^{(m)}_2 (P^{(m)}_1)^\sharp P'^{(m)}_2)^\flat \ldots)^\flat P'^{(m)}_{m-1})^\flat_{2,2} = \infty$$
\end{lem}

\begin{lem}
\label{lemma:two}
Let $1< m\leq n$. If for all $1\leq i < m$, $g_ig'_i$ has only factors $\boldsymbol{a_j}$ for $j\leq i$ and $g_mg'_m$ has factor $\boldsymbol{a_j}$ for some $j> m$ then:
$$(P^{(m)}_{n-1} (\ldots (P^{(m)}_2 (P^{(m)}_1)^\sharp P'^{(m)}_2)^\flat \ldots)^\flat P'^{(m)}_{n-1})^\flat_{2,2} = \infty$$
\end{lem}

\begin{lem}
\label{lemma:three}
Let $1< m\leq n$. If $P$ is a $3\times 3$ matrix such that $P_{2,2} = \infty$, then the matrix $P^{(m)}_{n} P P'^{(m)}_{n}$ has $\infty$ in one of the $(1,2)$, $(2,2)$, $(1,3)$ or $(2,3)$ entries. 
\end{lem}

Before proving these three lemmas,  we show how they are used to prove that $g$ is not a tractable witness, entailing \cref{prop:last}.

\begin{proof}[Proof of \cref{prop:last}]
Let $z$ be the smallest integer between $1$ and $n$ such that for all $1\leq i <z$, $g_ig'_i$ has only factors $\boldsymbol{a_j}$ for $j\leq i$, and $g_zg'_z$ has factor $\boldsymbol{a_j}$ for some $j> z$; and $\bot$ if such an integer does not exist. We are in one of the following cases:
\begin{itemize}
    \item $z = \bot$: In this case, for all $1\leq i \leq n$, $g_ig'_i$ has only factors $\boldsymbol{a_j}$ for $j\leq i$.
    By denoting
    $$P = (P^{(n)}_{n-1} (\ldots (P^{(n)}_2 (P^{(n)}_1)^\sharp P'^{(n)}_2)^\flat \ldots)^\flat P'^{(n)}_{n-1})^\flat$$
    we have, by Lemma~\ref{lemma:one} (applied to $m=n$), 
    $$P_{2,2} = \infty$$
    and by Lemma~\ref{lemma:three}, 
    $$M^{(n)} = P^{(n)}_{n} P P'^{(n)}_{n}$$
    has $\infty$ in one of the $(1,2)$, $(2,2)$, $(1,3)$ or $(2,3)$ entries. 
    \item $1 < z \leq n$: By denoting
    $$P = (P^{(z)}_{n-1} (\ldots (P^{(z)}_2 (P^{(z)}_1)^\sharp P'^{(z)}_2)^\flat \ldots)^\flat P'^{(z)}_{n-1})^\flat$$
    we have, by Lemma~\ref{lemma:two} (applied to $m=z$), 
    $$P_{2,2} = \infty$$
    and by Lemma~\ref{lemma:three}, 
    $$M^{(z)} = P^{(z)}_{n} P P'^{(z)}_{n}$$
    has $\infty$ in one of the $(1,2)$, $(2,2)$, $(1,3)$ or $(2,3)$ entries.  
    \item $z=1$: $g_1$ has factor $\boldsymbol{a_j}$ for some $j> 1$, and hence $M^{(1)} = (\infty)$.\qedhere
\end{itemize}
\end{proof}
\paragraph{Proof of \cref{lemma:one,lemma:three}.}
\begin{proof}[Proof of Lemma~\ref{lemma:one}.]
The proof is by induction on $m$.
If $m = 2$, and $g_1$ has only factors $\boldsymbol{a_1}$ then:
$$(P^{(2)}_1)^\sharp_{2,2} = \infty$$
Let $2<m \leq n$, and suppose that for all $1\leq i < m$, $g_ig'_i$ has only factors $\boldsymbol{a_j}$ for $j\leq i$. By induction, we have that 
$$(P^{(m-1)}_{m-2} (\ldots (P^{(m-1)}_2 (P^{(m-1)}_1)^\sharp P'^{(m-1)}_2)^\flat \ldots)^\flat P'^{(m-1)}_{m-2})^\flat_{2,2} = \infty$$
Since $\mathcal{B}_{n,m-1}$ and $\mathcal{B}_{n,m}$ are identical when restricted to letters $a_1, \ldots, a_{m-2}$, then:
$$(P^{(m)}_{m-2} (\ldots (P^{(m)}_2 (P^{(m)}_1)^\sharp P'^{(m)}_2)^\flat \ldots)^\flat P'^{(m)}_{m-2})^\flat_{2,2} = \infty$$
Finally, since $g_{m-1}g'_{m-1}$ has only factors $\boldsymbol{a_j}$ for $j\leq m-1$:
\begin{equation*}(P^{(m)}_{m-1} (\ldots (P^{(m)}_2 (P^{(m)}_1)^\sharp P'^{(m)}_2)^\flat \ldots)^\flat P'^{(m)}_{m-1})^\flat_{2,2} = \infty\qedhere
\end{equation*}
\end{proof}

\begin{proof}[Proof of Lemma~\ref{lemma:three}.]
If $g_{n}$ has factor $\boldsymbol{a_m}$, then $(P^{(m)}_{n})_{1,2} \neq -\infty$, otherwise $(P^{(m)}_{n})_{2,2} \neq -\infty$. 

If $g'_{n}$ does not have factor $\boldsymbol{a_m}$, or has at least one factor $\boldsymbol{a_m}$ that is later followed by a factor $\boldsymbol{a_j}$ for $j>m$, then $(P^{(m)}_{n})_{2,2} \neq -\infty$, otherwise, if no factor $\boldsymbol{a_m}$ is later followed by a factor $\boldsymbol{a_j}$ for $j>m$, $(P^{(m)}_{n})_{2,3} \neq -\infty$. 

In all the cases, the matrix $P^{(m)}_{n} P P'^{(m)}_{n}$ has $\infty$ in one of the $(1,2)$, $(2,2)$, $(1,3)$ or $(2,3)$ entries.
\end{proof}

\paragraph{Proof of Lemma~\ref{lemma:two}.}
Given a word $w$, we define $P^{(m)}_w$ as the $3\times 3$ matrix of the behaviour of the automaton $\mathcal{B}_{n,m}$ on the word $w$. We extend this definition to $\sharp \flat$-expressions on $\Sigma_n$:
\begin{itemize}
\item If $w$ is a word, then $P^{(m)}_w$ is defined above.
\item If $w = w_1w_2$ then $P^{(m)}_w = P^{(m)}_{w_1}P^{(m)}_{w_2}$.
\item If $w = u^\sharp$ then $P^{(m)}_w = (P^{(m)}_{u})^\sharp$ provided $P^{(m)}_{u}$ is path-idempotent, otherwise it is undefined.
\item If $w = u^\flat$ then $P^{(m)}_w = (P^{(m)}_{u})^\flat$ provided $P^{(m)}_{u}$ is path-idempotent, otherwise it is undefined.
\end{itemize}
Essentially, if $\pi(w) = (q,x,q, M)$, then $P_w^{(m)}$ is the restriction of $M$ to matrix for $B_{n,m}$.

We say that a $3\times 3$ matrix $P$ satisfies condition (C) if the four following conditions are satisfied: 
\begin{description}
    
\item[C1] $P_{2,1} \neq -\infty$, 
\item[C2] $P_{3,1} \neq -\infty$,
\item[C3] $P_{3,2} \neq -\infty$,
\item[C4] $P_{2,2} = \infty$.
\end{description}

\begin{lem}
\label{lemma:twoA}
Let $v$ and $w$ be two words and $u$ a $\sharp \flat$-expression on $\Sigma_n$ such that $(P^{(m)}_u)$ is defined, $(P^{(m)}_u)_{2,2} = \infty$ and $vw$ has a letter $a_j$ for some $j>m$ as factor. 
Then $(P^{(m)}_v P^{(m)}_u P^{(m)}_w)^\flat$ satisfies (C).
\end{lem}

\begin{lem}
\label{lemma:twoB}
Let $v$ and $w$ be two words and $u$ a $\sharp \flat$-expression on $\Sigma_n$ such that $(P^{(m)}_u)$ is defined and satisfies (C) then $(P^{(m)}_v P^{(m)}_u P^{(m)}_w)^\flat$ satisfies (C).
\end{lem}

Let us first show how we can use Lemmas~\ref{lemma:twoA} and~\ref{lemma:twoB} to conclude the proof. We prove by induction that for all $m \leq k < n$, 
$$(P^{(m)}_{k} (\ldots (P^{(m)}_2 (P^{(m)}_1)^\sharp P'^{(m)}_2)^\flat \ldots)^\flat P'^{(m)}_{k})^\flat$$
satisfies (C). If so, by applying it to $k=n$ and considering (C4), this concludes the proof.

The base case ($k=m$) is given by Lemma~\ref{lemma:twoA} applied to 
$$P^{(m)}_u = (P^{(m)}_{m-1} (\ldots (P^{(m)}_2 (P^{(m)}_1)^\sharp P'^{(m)}_2)^\flat \ldots)^\flat P'^{(m)}_{m-1})^\flat$$
$P^{(m)}_v = P^{(m)}_m$ and $P^{(m)}_w = P'^{(m)}_m$. $(P^{(m)}_u)_{2,2} = \infty$ by Lemma~\ref{lemma:one}.

The induction case is given by Lemma~\ref{lemma:twoB}. 

\paragraph{Proof of Lemmas~\ref{lemma:twoA} and~\ref{lemma:twoB}.}

First note the following properties:
\begin{description}
    
    \item[Property 1] for all $\sharp \flat$-expressions $w$, $(P^{(m)}_w)_{1,1} \neq -\infty$.
    \item[Property 2] for all $\sharp \flat$-expressions $w$, one (or both) of  $(P^{(m)}_w)_{2,2}$ or $(P^{(m)}_w)_{2,3}$ is not $-\infty$.
    \item[Property 3] for all $\sharp \flat$-expressions $w$, one (or both) of  $(P^{(m)}_w)_{3,2}$ or $(P^{(m)}_w)_{3,3}$ is not $-\infty$.
    \item[Property 4] for all words $w$ that has a letter $a_j$ for some $j>m$ as factor, $(P^{(m)}_w)_{2,1}$ and $(P^{(m)}_w)_{3,1}$ are not $-\infty$.
    \item[Property 5] for all words $w$ that has a letter $a_j$ for some $j>m$ as factor, $(P^{(m)}_w)_{2,2}$ and $(P^{(m)}_w)_{3,2}$ are not $-\infty$. (For entry $(2,2)$: Given a word $w$ with a letter $a_j$ for some $j>m$ as factor, (1) if $w$ does not have letter $a_m$ as factor then there is a path looping around $q_{m,2}$ on $w$, (2) if $w$ has a letter $a_m$ and all letters $a_m$ are followed later in the word by a letter $a_\ell$ for some $\ell > m$ then loop around $q_{m,2}$ until reading a $a_m$, then go to $q_{m,3}$, loop around $q_{m,3}$ until seeing a letter $a_\ell$ for some $\ell > m$ and go back to $q_{m,2}$. Repeat, (3) if $w$ has a letter $a_m$ and some are not followed later by a letter $a_\ell$ for some $\ell > m$, then stay in $q_{m,2}$ or go to $q_{m,3}$ as long as no $a_\ell$ for some $\ell > m$ is seen. When the first one is read, go to $q_{m,1}$ and on the very last $a_m$ go back to $q_{m,2}$.
    For entry $(3,2)$: Given a word $w$ with a letter $a_j$ for some $j>m$ as factor, (1) if $w$ does not have letter $a_m$ as factor then go to $q_{m,2}$ when reading the first $a_j$ and loop in $q_{m,2}$ on the rest of the word, (2) if $w$ has a letter $a_m$ and all letters $a_m$ are followed later in the word by a letter $a_\ell$ for some $\ell > m$ then loop around $q_{m,3}$ until reading $a_\ell$ for some $\ell > m$ and go to $q_{m,2}$, loop around $q_{m,2}$ until seeing a $a_m$, then go to $q_{m,3}$, loop around $q_{m,3}$ until seeing a letter $a_\ell$ for some $\ell > m$ and go back to $q_{m,2}$. Repeat, (3) if $w$ has a letter $a_m$ and some are not followed later by a letter $a_\ell$ for some $\ell > m$, then stay in $q_{m,3}$ or go to $q_{m,2}$ as long as no $a_\ell$ for some $\ell > m$ is seen. When the first one is read, go to $q_{m,1}$ and on the very last $a_m$ go back to $q_{m,2}$.
    ) 
    \item[Property 6] for all $\sharp \flat$-expressions $w$, one (or both) of  $(P^{(m)}_w)_{1,2}$ or $(P^{(m)}_w)_{2,2}$ is not $-\infty$.

\end{description}

\begin{proof}[Proof of Lemma~\ref{lemma:twoA}]
By path-idempotency, for (C1), (C2) and (C3), it is sufficient to prove that $P^{(m)}_v P^{(m)}_u P^{(m)}_w$ satisfies (C1), (C2) and (C3). 
\begin{description}
    
\item[C1 ($P_{2,1} \neq -\infty$)]  If $v$ has factor $a_j$ for some $j>m$, then $(P^{(m)}_v)_{2,1}$ is not $-\infty$ (Property 4). Additionally, $(P^{(m)}_u)_{1,1}$ and $(P^{(m)}_w)_{1,1}$ are not $-\infty$ (Property 1). \\
If $w$ has factor $a_j$ for some $j>m$, then neither $(P^{(m)}_w)_{2,1}$ or $(P^{(m)}_w)_{3,1}$ is $-\infty$ (Property 4). Additionally, $P^{(m)}_v P^{(m)}_u = P^{(m)}_{vu}$ and one of $(P^{(m)}_{vu})_{2,2}$ or $(P^{(m)}_{vu})_{2,3}$ is not $-\infty$ (Property 2). \\
In all cases, $(P^{(m)}_v P^{(m)}_u P^{(m)}_w)_{2,1}$ is not $-\infty$.

\item[C2 ($P_{3,1} \neq -\infty$)]  If $v$ has factor $a_j$ for some $j>m$, then $(P^{(m)}_v)_{3,1}$ is not $-\infty$ (Property 4). Additionally, $(P^{(m)}_u)_{1,1}$ and $(P^{(m)}_w)_{1,1}$ are not $-\infty$ (Property 1). \\
If $w$ has factor $a_j$ for some $j>m$, then neither $(P^{(m)}_w)_{2,1}$ or $(P^{(m)}_w)_{3,1}$ is $-\infty$ (Property 4). Additionally, $P^{(m)}_v P^{(m)}_u = P^{(m)}_{vu}$ and one of $(P^{(m)}_{vu})_{3,2}$ or $(P^{(m)}_{vu})_{3,3}$ is not $-\infty$ (Property 3).\\ 
In all cases, $(P^{(m)}_v P^{(m)}_u P^{(m)}_w)_{3,1}$ is not $-\infty$.

\item[C3 ($P_{3,2} \neq -\infty$)] If $w$ has factor $a_j$ for some $j>m$, then both $(P^{(m)}_w)_{2,2}$ and $(P^{(m)}_w)_{3,2}$ are not $-\infty$ (Property 5). Additionally, $P^{(m)}_v P^{(m)}_u = P^{(m)}_{vu}$ and one of $(P^{(m)}_{vu})_{3,2}$ or $(P^{(m)}_{vu})_{3,3}$ is not $-\infty$ (Property 3). \\
If $v$ has factor $a_j$ for some $j>m$, then both $(P^{(m)}_w)_{3,1}$ and $(P^{(m)}_w)_{3,2}$ are not $-\infty$ (Properties 4 and 5). Additionally, $P^{(m)}_u P^{(m)}_w = P^{(m)}_{uw}$ and one of $(P^{(m)}_{uw})_{1,2}$ or $(P^{(m)}_{uw})_{2,2}$ is not $-\infty$ (Property 6). \\
In all cases, $(P^{(m)}_v P^{(m)}_u P^{(m)}_w)_{3,2}$ is not $-\infty$.

\item[C4 ($P_{2,2} = \infty$)] We need to prove that:
$$\overline{P^{(m)}_v P^{(m)}_u P^{(m)}_w} \langle (P^{(m)}_v P^{(m)}_u P^{(m)}_w)^3 \rangle \overline{P^{(m)}_v P^{(m)}_u P^{(m)}_w}$$
satisfies (C4).
First, $(\overline{P^{(m)}_v P^{(m)}_u P^{(m)}_w})_{2,2}$ is not $-\infty$ since $vuw$ has a factor $a_j$ for some $j>m$ (Property 5). Hence, it remains to prove that $((P^{(m)}_v P^{(m)}_u P^{(m)}_w)^3)_{2,2}$ is $\infty$. We have one of $(P^{(m)}_v P^{(m)}_u)_{2,2}$ or $(P^{(m)}_v P^{(m)}_u)_{2,3}$ is not $-\infty$ (Property 2). Then both $(P^{(m)}_w P^{(m)}_v)_{2,2}$ and $(P^{(m)}_w P^{(m)}_v)_{3,2}$ are not $-\infty$ (Property 5). By hypothesis, we have $(P^{(m)}_u)_{2,2} = \infty$. And finally, $(P^{(m)}_{wvuw})_{2,2}$ is not $-\infty$ (Property 5). Hence, $\langle (P^{(m)}_v P^{(m)}_u P^{(m)}_w)^3 \rangle_{2,2} = \infty$.\qedhere

\end{description}
\end{proof}

\begin{proof}[Proof of Lemma~\ref{lemma:twoB}]
By path-idempotency, for (C1), (C2) and (C3), it is sufficient to prove that $P^{(m)}_v P^{(m)}_u P^{(m)}_w$ satisfies (C1), (C2) and (C3).

\begin{description}
    \item[C1 ($P_{2,1} \neq -\infty$)] One of $(P^{(m)}_v)_{2,2}$ or $(P^{(m)}_v)_{2,3}$ is not $-\infty$ (Property 2). Additionally, $(P^{(m)}_u)_{2,1}$ and $(P^{(m)}_u)_{3,1}$ are not $-\infty$ by (C1) and (C2), and $(P^{(m)}_w)_{1,1}$ is not $-\infty$ (Property 1). Hence $(P^{(m)}_v P^{(m)}_u P^{(m)}_w)_{2,1}$ is not $-\infty$.

\item[C2 ($P_{3,1} \neq -\infty$)]One of $(P^{(m)}_v)_{3,2}$ or $(P^{(m)}_v)_{3,3}$ is not $-\infty$ (Property 3). Additionally, $(P^{(m)}_u)_{2,1}$ and $(P^{(m)}_u)_{3,1}$ are not $-\infty$ by (C1) and (C2), and $(P^{(m)}_w)_{1,1}$ is not $-\infty$ (Property 1). Hence $(P^{(m)}_v P^{(m)}_u P^{(m)}_w)_{2,1}$ is not $-\infty$.

\item[C3 ($P_{3,2} \neq -\infty$)]One of $(P^{(m)}_v)_{3,2}$ or $(P^{(m)}_v)_{3,3}$ is not $-\infty$ (Property 3). Additionally, $(P^{(m)}_u)_{2,1}$, $(P^{(m)}_u)_{2,2}$, $(P^{(m)}_u)_{3,1}$ and $(P^{(m)}_u)_{3,2}$ are not $-\infty$ by (C). Finally, one of $(P^{(m)}_{w})_{1,2}$ or $(P^{(m)}_{w})_{2,2}$ is not $-\infty$ (Property 6). Hence, overall, $(P^{(m)}_v P^{(m)}_u P^{(m)}_w)_{3,2}$ is not $-\infty$.

\item[C4 ($P_{2,2} = \infty$)] We need to prove that:
$$\overline{P^{(m)}_v P^{(m)}_u P^{(m)}_w} \langle (P^{(m)}_v P^{(m)}_u P^{(m)}_w)^3 \rangle \overline{P^{(m)}_v P^{(m)}_u P^{(m)}_w}$$
satisfies (C4).
First, let's remark that for all words $x$ and $y$, $(P^{(m)}_x P^{(m)}_u P^{(m)}_y)_{2,2}$ is not $-\infty$. Indeed, one of $(P^{(m)}_x)_{2,2}$ or $(P^{(m)}_x)_{2,3}$ is not $-\infty$ (Property 2). Similarly, one of $(P^{(m)}_y)_{1,2}$ or $(P^{(m)}_x)_{2,2}$ is not $-\infty$ (Property 6). And additionally, all of $(P^{(m)}_u)_{2,1}$, $(P^{(m)}_u)_{2,2}$,  $(P^{(m)}_u)_{2,1}$,  $(P^{(m)}_u)_{3,2}$ are not $-\infty$ (condition (C)). Hence, $(P^{(m)}_x P^{(m)}_u P^{(m)}_y)_{2,2}$ is not $-\infty$.
That gives that: $(\overline{P^{(m)}_v P^{(m)}_u P^{(m)}_w})_{2,2}$ is not $-\infty$, leaving us to prove that $((P^{(m)}_v P^{(m)}_u P^{(m)}_w)^3)_{2,2}$ is $\infty$. Still using the remark, we have that both $(P^{(m)}_v P^{(m)}_u P^{(m)}_{wv})_{2,2}$ and $(P^{(m)}_{wv} P^{(m)}_u P^{(m)}_w)_{2,2}$ are not $-\infty$, which concludes the proof, using finally the hypothesis that $(P^{(m)}_u)_{2,2} = \infty$. \qedhere
\end{description}
\end{proof}

\section{Conclusion}
In this paper, we develop new techniques -- in particular a new flattening operation -- to describe the behaviour of max-plus automata. It would be interesting to see if such insight can be applied to other problems, particularly for min-plus automata. We also construct a series of examples of max-plus automata requiring tractable witnesses with an increasing number of nested flattening operations\footnote{We thank Isma\"el Jecker for interesting discussions on this.}.

\bibliographystyle{alphaurl}
\bibliography{ref}

\newcommand{\etalchar}[1]{$^{#1}$}
\begin{thebibliography}{KMO{\etalchar{+}}13}

\bibitem[ABK11]{AlmagorBK11}
Shaull Almagor, Udi Boker, and Orna Kupferman.
\newblock What's decidable about weighted automata?
\newblock In {\em Automated Technology for Verification and Analysis, 9th
  International Symposium, {ATVA} 2011}, volume 6996 of {\em Lecture Notes in
  Computer Science}, pages 482--491. Springer, 2011.
\newblock \href {https://doi.org/10.1007/978-3-642-24372-1_37}
  {\path{doi:10.1007/978-3-642-24372-1_37}}.

\bibitem[ABK22]{AlmagorBK22}
Shaull Almagor, Udi Boker, and Orna Kupferman.
\newblock What's decidable about weighted automata?
\newblock {\em Inf. Comput.}, 282:104651, 2022.
\newblock Journal version of~\cite{AlmagorBK11}.
\newblock \href {https://doi.org/10.1016/j.ic.2020.104651}
  {\path{doi:10.1016/j.ic.2020.104651}}.

\bibitem[CD13]{ColcombetD13}
Thomas Colcombet and Laure Daviaud.
\newblock Approximate comparison of distance automata.
\newblock In {\em 30th International Symposium on Theoretical Aspects of
  Computer Science, {STACS} 2013}, volume~20 of {\em LIPIcs}, pages 574--585.
  Schloss Dagstuhl - Leibniz-Zentrum f{\"{u}}r Informatik, 2013.
\newblock \href {https://doi.org/10.4230/LIPIcs.STACS.2013.574}
  {\path{doi:10.4230/LIPIcs.STACS.2013.574}}.

\bibitem[CDH10]{ChatterjeeDH10}
Krishnendu Chatterjee, Laurent Doyen, and Thomas~A. Henzinger.
\newblock Quantitative languages.
\newblock {\em {ACM} Trans. Comput. Log.}, 11(4):23:1--23:38, 2010.
\newblock \href {https://doi.org/10.1145/1805950.1805953}
  {\path{doi:10.1145/1805950.1805953}}.

\bibitem[CDZ14]{ColcombetDZ14}
Thomas Colcombet, Laure Daviaud, and Florian Zuleger.
\newblock Size-change abstraction and max-plus automata.
\newblock In {\em Mathematical Foundations of Computer Science 2014 - 39th
  International Symposium, {MFCS} 2014}, volume 8634 of {\em Lecture Notes in
  Computer Science}, pages 208--219. Springer, 2014.
\newblock \href {https://doi.org/10.1007/978-3-662-44522-8_18}
  {\path{doi:10.1007/978-3-662-44522-8_18}}.

\bibitem[CKMP20]{ChistikovKMP20}
Dmitry Chistikov, Stefan Kiefer, Andrzej~S. Murawski, and David Purser.
\newblock The big-{O} problem for labelled markov chains and weighted automata.
\newblock In {\em 31st International Conference on Concurrency Theory, {CONCUR}
  2020}, volume 171 of {\em LIPIcs}, pages 41:1--41:19. Schloss Dagstuhl -
  Leibniz-Zentrum f{\"{u}}r Informatik, 2020.
\newblock Conference version of \cite{ChistikovKMP22}.
\newblock \href {https://doi.org/10.4230/LIPIcs.CONCUR.2020.41}
  {\path{doi:10.4230/LIPIcs.CONCUR.2020.41}}.

\bibitem[CKMP22]{ChistikovKMP22}
Dmitry Chistikov, Stefan Kiefer, Andrzej~S. Murawski, and David Purser.
\newblock The big-{O} problem.
\newblock {\em Log. Methods Comput. Sci.}, 18(1), 2022.
\newblock Journal version of \cite{ChistikovKMP20}.
\newblock \href {https://doi.org/10.46298/lmcs-18(1:40)2022}
  {\path{doi:10.46298/lmcs-18(1:40)2022}}.

\bibitem[CLM{\etalchar{+}}22]{CzerwinskiLMPW22}
Wojciech Czerwinski, Engel Lefaucheux, Filip Mazowiecki, David Purser, and
  Markus~A. Whiteland.
\newblock The boundedness and zero isolation problems for weighted automata
  over nonnegative rationals.
\newblock In {\em {LICS} '22: 37th Annual {ACM/IEEE} Symposium on Logic in
  Computer Science, Haifa, Israel, August 2 - 5, 2022}, pages 15:1--15:13.
  {ACM}, 2022.
\newblock \href {https://doi.org/10.1145/3531130.3533336}
  {\path{doi:10.1145/3531130.3533336}}.

\bibitem[Col07]{Colcombet07}
Thomas Colcombet.
\newblock Factorisation forests for infinite words.
\newblock In {\em Fundamentals of Computation Theory, 16th International
  Symposium, {FCT} 2007}, volume 4639 of {\em Lecture Notes in Computer
  Science}, pages 226--237. Springer, 2007.
\newblock \href {https://doi.org/10.1007/978-3-540-74240-1_20}
  {\path{doi:10.1007/978-3-540-74240-1_20}}.

\bibitem[Col09]{Colcombet09}
Thomas Colcombet.
\newblock The theory of stabilisation monoids and regular cost functions.
\newblock In {\em Automata, Languages and Programming, 36th Internatilonal
  Colloquium, {ICALP} 2009, Rhodes, Greece, July 5-12, 2009, Proceedings, Part
  {II}}, volume 5556 of {\em Lecture Notes in Computer Science}, pages
  139--150. Springer, 2009.
\newblock \href {https://doi.org/10.1007/978-3-642-02930-1_12}
  {\path{doi:10.1007/978-3-642-02930-1_12}}.

\bibitem[DJL{\etalchar{+}}21]{DaviaudJLMP021}
Laure Daviaud, Marcin Jurdzinski, Ranko Lazic, Filip Mazowiecki, Guillermo~A.
  P{\'{e}}rez, and James Worrell.
\newblock When are emptiness and containment decidable for probabilistic
  automata?
\newblock {\em J. Comput. Syst. Sci.}, 119:78--96, 2021.
\newblock \href {https://doi.org/10.1016/j.jcss.2021.01.006}
  {\path{doi:10.1016/j.jcss.2021.01.006}}.

\bibitem[DP23]{DaviaudP23}
Laure Daviaud and David Purser.
\newblock The big-{O} problem for max-plus automata is decidable
  ({PSPACE}-complete).
\newblock In {\em 38th Annual {ACM/IEEE} Symposium on Logic in Computer
  Science, {LICS} 2023, Boston, MA, USA, June 26-29, 2023}, pages 1--13.
  {IEEE}, 2023.
\newblock \href {https://doi.org/10.1109/LICS56636.2023.10175798}
  {\path{doi:10.1109/LICS56636.2023.10175798}}.

\bibitem[FGKO15]{FijalkowGKO15}
Nathana{\"{e}}l Fijalkow, Hugo Gimbert, Edon Kelmendi, and Youssouf Oualhadj.
\newblock Deciding the value 1 problem for probabilistic leaktight automata.
\newblock {\em Log. Methods Comput. Sci.}, 11(2), 2015.
\newblock \href {https://doi.org/10.2168/LMCS-11(2:12)2015}
  {\path{doi:10.2168/LMCS-11(2:12)2015}}.

\bibitem[Has91]{Hashiguchi91a}
Kosaburo Hashiguchi.
\newblock Algorithms for determining relative inclusion star height and
  inclusion star height.
\newblock {\em Theor. Comput. Sci.}, 91(1):85--100, 1991.
\newblock \href {https://doi.org/10.1016/0304-3975(91)90269-8}
  {\path{doi:10.1016/0304-3975(91)90269-8}}.

\bibitem[KL09]{KirstenL09}
Daniel Kirsten and Sylvain Lombardy.
\newblock Deciding unambiguity and sequentiality of polynomially ambiguous
  min-plus automata.
\newblock In {\em 26th International Symposium on Theoretical Aspects of
  Computer Science, {STACS} 2009, February 26-28, 2009, Freiburg, Germany,
  Proceedings}, volume~3 of {\em LIPIcs}, pages 589--600. Schloss Dagstuhl -
  Leibniz-Zentrum f{\"{u}}r Informatik, Germany, 2009.
\newblock \href {https://doi.org/10.4230/LIPIcs.STACS.2009.1850}
  {\path{doi:10.4230/LIPIcs.STACS.2009.1850}}.

\bibitem[KMO{\etalchar{+}}13]{kiefer13}
Stefan Kiefer, Andrzej~S. Murawski, Jo{\"{e}}l Ouaknine, Bj{\"{o}}rn Wachter,
  and James Worrell.
\newblock On the complexity of equivalence and minimisation for {Q}-weighted
  automata.
\newblock {\em Log. Methods Comput. Sci.}, 9(1), 2013.
\newblock \href {https://doi.org/10.2168/LMCS-9(1:8)2013}
  {\path{doi:10.2168/LMCS-9(1:8)2013}}.

\bibitem[Kro94]{Krob94}
Daniel Krob.
\newblock The equality problem for rational series with multiplicities in the
  tropical semiring is undecidable.
\newblock {\em Int. J. Algebra Comput.}, 4(3):405--426, 1994.
\newblock \href {https://doi.org/10.1142/S0218196794000063}
  {\path{doi:10.1142/S0218196794000063}}.

\bibitem[Kuf08]{Kufleitner08}
Manfred Kufleitner.
\newblock The height of factorization forests.
\newblock In {\em Mathematical Foundations of Computer Science 2008, 33rd
  International Symposium, {MFCS} 2008}, volume 5162, pages 443--454. Springer,
  2008.
\newblock \href {https://doi.org/10.1007/978-3-540-85238-4_36}
  {\path{doi:10.1007/978-3-540-85238-4_36}}.

\bibitem[Leu88]{leung1988topological}
Hing Leung.
\newblock On the topological structure of a finitely generated semigroup of
  matrices.
\newblock In {\em Semigroup Forum}, volume~37, pages 273--287. Springer, 1988.
\newblock \href {https://doi.org/10.1007/BF02573140}
  {\path{doi:10.1007/BF02573140}}.

\bibitem[MPR02]{MohriPR02}
Mehryar Mohri, Fernando Pereira, and Michael Riley.
\newblock Weighted finite-state transducers in speech recognition.
\newblock {\em Comput. Speech Lang.}, 16(1):69--88, 2002.
\newblock \href {https://doi.org/10.1006/csla.2001.0184}
  {\path{doi:10.1006/csla.2001.0184}}.

\bibitem[MS72]{MeyerS72}
Albert~R. Meyer and Larry~J. Stockmeyer.
\newblock The equivalence problem for regular expressions with squaring
  requires exponential space.
\newblock In {\em 13th Annual Symposium on Switching and Automata Theory,
  College Park, Maryland, USA, October 25-27, 1972}, pages 125--129. {IEEE}
  Computer Society, 1972.
\newblock \href {https://doi.org/10.1109/SWAT.1972.29}
  {\path{doi:10.1109/SWAT.1972.29}}.

\bibitem[Paz14]{paz2014introduction}
Azaria Paz.
\newblock {\em Introduction to probabilistic automata}.
\newblock Academic Press, 2014.
\newblock \href {https://doi.org/10.1016/C2013-0-11297-4}
  {\path{doi:10.1016/C2013-0-11297-4}}.

\bibitem[Sch61]{Schutzenberger61b}
Marcel~Paul Sch{\"{u}}tzenberger.
\newblock On the definition of a family of automata.
\newblock {\em Inf. Control.}, 4(2-3):245--270, 1961.
\newblock \href {https://doi.org/10.1016/S0019-9958(61)80020-X}
  {\path{doi:10.1016/S0019-9958(61)80020-X}}.

\bibitem[Sim90]{Simon90}
Imre Simon.
\newblock Factorization forests of finite height.
\newblock {\em Theor. Comput. Sci.}, 72(1):65--94, 1990.
\newblock \href {https://doi.org/10.1016/0304-3975(90)90047-L}
  {\path{doi:10.1016/0304-3975(90)90047-L}}.

\bibitem[Sim94]{Simon94}
Imre Simon.
\newblock On semigroups of matrices over the tropical semiring.
\newblock {\em {RAIRO} Theor. Informatics Appl.}, 28(3-4):277--294, 1994.
\newblock \href {https://doi.org/10.1051/ita/1994283-402771}
  {\path{doi:10.1051/ita/1994283-402771}}.

\bibitem[SM73]{StockmeyerM73}
Larry~J. Stockmeyer and Albert~R. Meyer.
\newblock Word problems requiring exponential time: Preliminary report.
\newblock In {\em Proceedings of the 5th Annual {ACM} Symposium on Theory of
  Computing, 1973}, pages 1--9. {ACM}, 1973.
\newblock \href {https://doi.org/10.1145/800125.804029}
  {\path{doi:10.1145/800125.804029}}.

\bibitem[Var85]{Vardi85}
Moshe~Y. Vardi.
\newblock Automatic verification of probabilistic concurrent finite-state
  programs.
\newblock In {\em 26th Annual Symposium on Foundations of Computer Science,
  Portland, Oregon, USA, 21-23 October 1985}, pages 327--338. {IEEE} Computer
  Society, 1985.
\newblock \href {https://doi.org/10.1109/SFCS.1985.12}
  {\path{doi:10.1109/SFCS.1985.12}}.

\end{thebibliography}

\end{document}